\documentclass[%
 reprint,
 amsmath,amssymb,
 aps,onecolumn,10pt]{revtex4-1}
\usepackage{subfig}
\usepackage{graphicx}
\usepackage{float}
\usepackage{dcolumn}
\usepackage{bm}
\usepackage{sistyle}
\usepackage{braket}
\usepackage{lipsum}
\usepackage{color}
\usepackage{titlesec}
\usepackage[]{subfig}
\begin{document}
\title{Optical properties of exciton in two-dimensional transition metal dichalcogenide nanobubbles}

\author{Adlen Smiri}
\affiliation{Facult\'e des Sciences de Bizerte, Laboratoire de Physique des Mat\'eriaux: Structure et Propri\'et\'es,\\
Universit\'e de Carthage, 7021 Jarzouna, Tunisia}%

 \author{Thierry Amand}
\affiliation{%
LPCNO, Universit\'e F\'ed\'erale de Toulouse Midi-Pyr\'en\'ees,\\
INSA-CNRS-UPS, 135 Av. de Rangueil, 31077 Toulouse, France
}%
\author{Sihem Jaziri}
\affiliation{Facult\'e des Sciences de Bizerte, Laboratoire de Physique des Mat\'eriaux: Structure et Propri\'et\'es,\\
Universit\'e  de Carthage, 7021 Jarzouna, Tunisia
}%
\affiliation{
Facult\'e des Sciences de Tunis, Laboratoire de Physique de la Mati\'ere Condens\'ee,\\
D\'epartement de Physique, Universit\'e  Tunis el Manar, Campus Universitaire 2092 Tunis, Tunisia
}%
%to 
%\collaboration{CLEO Collaboration}
% $m_{d(a)}$
%\date{\today}
\begin{abstract}
Strain in two-dimensional (2D) transition metal dichalcogenide (TMD) has led to localized states with exciting optical properties, in particular in view of designing one photon sources. The naturally formed of the MoS$_2$ monolayer deposed on hBN substrate leads to a reduction of the bandgap in the strained region creating a nanobubble. The photogenerated particles are thus confined in the strain-induced potential. Using numerical diagonalization, we simulate the spectra of the confined exciton states, their oscillator strengths and radiative lifetimes. We show that a single state of the confined exciton is optically active, which suggests that the MoS$_2$/hBN nanobubble is a good candidate for the realisation of single-photon sources. Furthermore, the exciton binding energy, oscillator strength and radiative lifetime are enhanced due to the confinement effect.
\end{abstract}
\pacs{Valid PACS appear here}
\maketitle
\section{Introduction}
Thanks to their uniqueness, the optical properties of two-dimensional (2D) transition metal dichalcogenide semiconductors (TMDs) still remains a matter of interest to the researchers~\cite{Geim,shepard,falko,lagarde,mak,mak2,ramasubramaniam,shepard,ayari,hanan,hichri,hichri2,chernikov,kylanpaa,graphite,olsen,hill,Robert}. The peculiar structural and electronic properties~\cite{falko,liu} of 2D-TMDs offer a various types of exciton with large binding energies. Indeed, the A- and B-bright excitons and the dark excitons~\cite{Robert} are observed in these materials due to the large spin-orbit coupling caused by the lack of space inversion symmetry~\cite{falko,liu}. Besides, there are a brigth and dark exciton complexes such the trions and biexcitons~\cite{kylanpaa,hichri,hichri2,falko}. The exciton characteristics, such as their binding energy, dratscally depend on the dielectric environment since these monolayers are on the atomic scale~\cite{lin,kylanpaa,olsen,ayari,hanan,hichri,hichri2}. In addition to the free exciton complexes, localized excitons have left traces in the photolumenescence (PL) spectrum~\citep{lagarde,mak,mak2}. Different proposals on the origin of localized excitons, which is still not fully understood, have been published~\cite{ayari,shepard}. For instance, we note the impurity or structural defect bound excitons and excitons trapped in local potential created by nanobubbles~\cite{srivastava,shepard,falko,ayari}. Recently, great interest has been given to localized excitons in 2D-TMDs. This focus of interest is not only motivated by the investigation of the origin of these excitons, but also by the potential applications taking advantage of their characteristics. Indeed, the localization of excitons can lead to the production of single photon emitters~\cite{chakraborty,koperski,he,tonndor} which is a main objective in quantum optics. In addition, localized excitons are characterized by their long radiative life which is required for photovoltaics.\\

Strain in 2D-TMDs has lead to an exciting localized optical properties~\cite{Geim,shepard,habib,andres,Artur,Artur2,carmen,david,Hong}. The 2D-TMD materials, because of theirs pecular mechanical properties, represent an ideal condidates for strain engeneering~\cite{Rafael,habib,andres,keliang,plechinger}. Indeed, these 2D materials can withstand extreme non-homogeneous deformations before rupture~\cite{andres}. For all 2D-TMDs under strain, there is a common important property which is the reduction of the bandgap  locally on the strained region~\cite{david,andres,Artur}. A quantum confinement then appears which confine photogenerated particles, the so-called strain-induced confinement~\cite{chirolli,andres,Artur}.  Based on this picture, several experimental strategies have been considered in order to create strain regions in 2D-TMDs that work as quantum emitters~\cite{david,Artur,carmen,shepard}. On the other hand, unintentional strained regions creating nanobubble were observed in MoS$_2$ monolayer (ML)~\cite{carmesin,chirolli,Geim}. For MoS$_2$ over a given substrate, the nanobubbles have an universal shape characterized by the height-over-radius aspect ratio~\cite{carmesin,chirolli,Geim}. The formation of bubbles is attributed to the competition between vdW forces and elastic energy \cite{Geim,chirolli}. Typical sizes of (nano)bubbles range from sub-10 nm to sub-micron \cite{chirolli,Geim}. Theoretically, in Ref. \cite{chirolli}, nanobubbles of up to 10 nm in size, were simulated. In this latter work, a sizeable reduction of the local bandgap occurs in the nanobubble, that acts as a potential well for conduction band carriers. In Ref. \cite{chirolli} Chirolli et \textit{al.} studied the case of a single particle confinement in the nanobubble. Until now, to our knowledge, there are no theoretical work in which a systematic investigation of the characteristics of the confined exciton in the nanobubbles has been undertaken.\\
 
In this work, we theoretically investigate the optical properties of exciton located in nanobubbles that naturally arise in MoS$_2$ ML deposited over a hBN substrate. We determine the exciton binding energies, oscillator strengths and radiative lifetime of the exciton dependence on the nanobubble size.
\section{Model}
\subsection{ The Schr\"odinger equation}
\begin{figure} %[h]
  \begin{center}
      \subfloat[]{
      \includegraphics[width=0.8\textwidth]{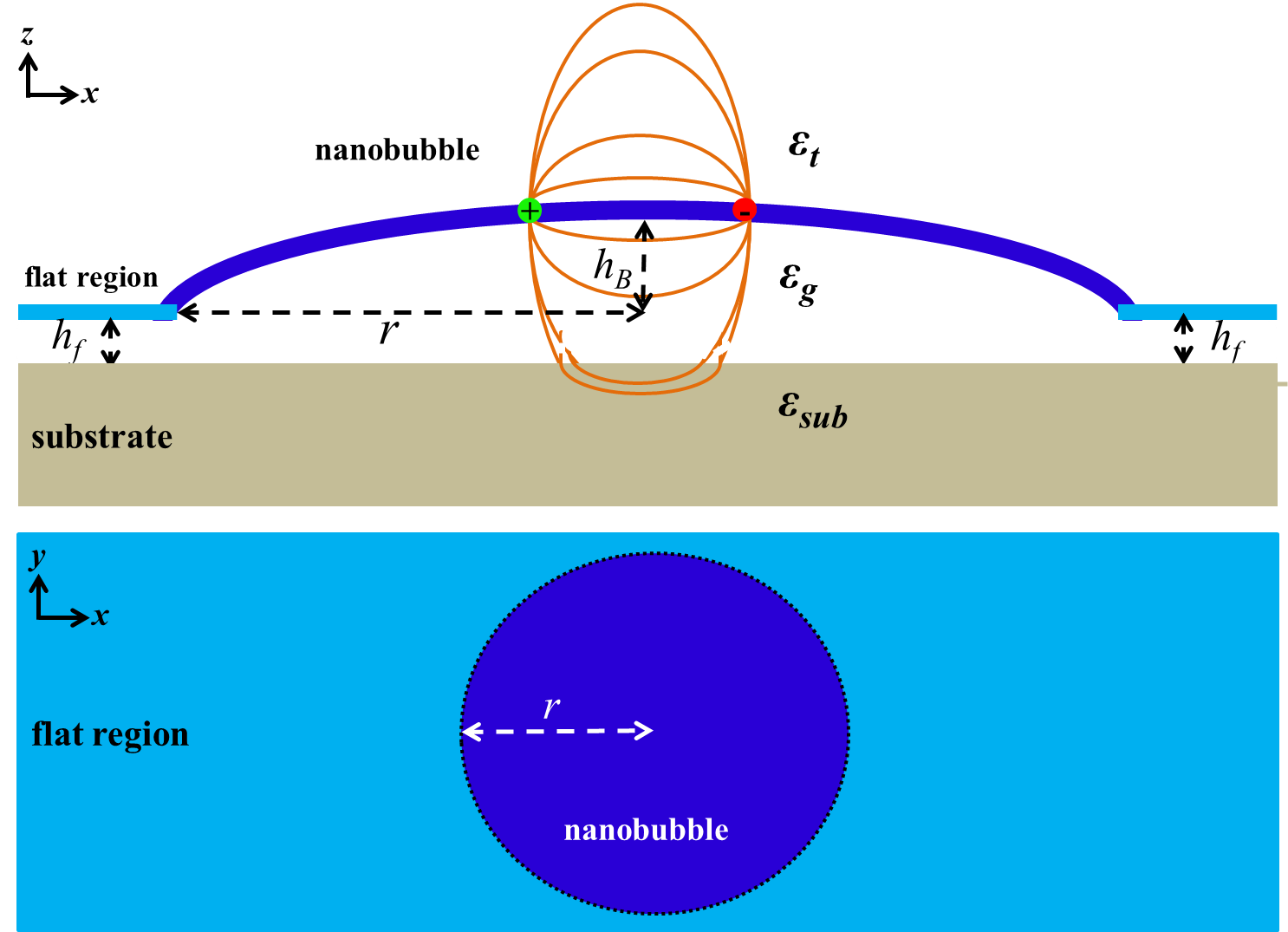}
                         }
   \captionsetup{format=plain,labelsep=period}                                                      
   \caption{ Schematic view of the circular nanobubble in $xz-$plane (the upper panel) and in $xy-$plane (the lower panel). $r$ and $h_B$ denote the radius and the maximum height of the nanobubble, respectively. $h_f$ is the interlayer gap width in the flat region. The center of the nanobubble is taken here as the origin of coordinates.}
    \label{fig(0)}
  \end{center}
\end{figure}

 The 2D exciton, interacting electron and hole, is confined in a nanobubble surface of MoS$_2$ ML deposed on hBN substrate. According to the Ref.~\cite{chirolli}, the nanobubble can be considered as a disklike quantum dot of radius $r$. The origin of coordinates in the crystal is taken at the disk center as seen in figure~(\ref{fig(0)}). This system is described by the following envelope Schr\"odinger equation in effective mass approximation, 
\begin{align}
\label{eq1}
\nonumber
& \lbrace-\frac{\hbar^2}{2m_e}\nabla_{r_e}^2-\frac{\hbar^2}{2m_h}\nabla_{r_h}^2-V_{SC}(|\textbf{r}_e-\textbf{r}_h|)\\
& +V_{B,e}(\textbf{r}_e)+V_{B,h}(\textbf{r}_h)\rbrace\Psi(\textbf{r}_e,\textbf{r}_h)=(E_{cf}-E_g)\Psi(\textbf{r}_e,\textbf{r}_h)
\end{align}
where $\bm{r}_e$ and $\bm{r}_h$ are the electron and hole coordinates, respectively and $V_{B,i}(r_i)(i=e,h)$ and $V_{SC}(|\bm{r_e}-\bm{r_h}|)$ denote the confinement and screened-Coulomb potentials, respectively. Here, $m_e$ and $m_h$ are the electron and hole effective masses of MoS$_2$ ML, respectively. While an agreement between theory and experiment on the value of $m_h=0.4m_0$~\cite{falko}, the experiment measures a $m_e$ of $0.7m_0$ which is two times greater than the theoretical value predicted by current DFT(-GW) calculations~\cite{pisoni}. $m_0$ is the free-electron mass. $E_{cf}$ is the confined exciton energy. $E_g=E_g(h_f)$ denotes the flat region band gap energy of MoS$_2$ ML which depends on the interlayer distance $h_f$. According to the Ref.~\cite{florian}, this latter was found equal to $h_f=5$ $\angstrom$ for the flat region of MoS$_2$/hBN system which gives $E_g=2.32$ eV. In the following, we analyse the different contributions to the potential in equation (1).
%%%%%%%%%%%%%%%%%%%%%%%%%%%%%%%%%%%%%%%%%%%%%%%%%%%%%%%%%%%%%%%%%%%%%%%%%%%%%%%%%%%%%%%%%%%%%%%%%%%%%%%%%%%%%%%%%%%%%%%%%%
%%%%%%%%%%%%%%%%%%%%%%%%%%%%%%%%%%%%%%%%%%%%%%%%%%%%%%%%%%%%%%%%%%%%%%%%%%%%%%%%
\subsubsection{Confinement potential}
The terms $V_{B,i}(r_i)(i=e,h)$ describe a reduction of the band gap in a local strained region of MoS$_2$ ML, i.e. the nanobubble, which leads to quantum confinement of particles~\cite{chirolli,Geim}. According to the Ref. \cite{chirolli}, the electron is the only particle that is affected by this confinement. In fact, the local reduction of the band gap is due to the locally reduced conduction band minima while the valence band maxima remains almost the same as the one of the flat region~\cite{chirolli}. Therefore, we consider in our work that $V_{B,h}(r_h)=0$. The confinement potential $V_{B,e}(r_e)$ can be modelled as~\cite{chirolli},
%by $V_e$ energy 
 \begin{equation}
 \label{eq2}
 V_{B,e}(r_e)= V_{B}(r_e)=-\Delta_0\bm{\xi}(r-r_e)
\end{equation}
  where, $\bm{\xi}(x)$ is the Heaviside function and $\Delta_0$ denotes the band edge difference of the flat region and the nanobubble. This latter can be written as $\Delta_0=\Delta_0^{str}-\Delta_0^{env}$ where $\Delta_0^{str}$ and $\Delta_0^{env}$ are the band gap local modifications induced by the strain and the dielectric environment, respectively. In contrast to the strain effect, the dielectric confinement tends to increase the band gap~\cite{chirolli,carmesin}. Therefore, there is a competition between both effects where a dominance of the strain effect is needed for the  quantum confinement to be effective.\\

The strain induced change in the band gap is given by $\displaystyle \Delta_0^{str}=\gamma \varpi_{max}$, where $\gamma\simeq6.4$ eV  is a proportionality constant of MoS$_2$ ML~\cite{chirolli}. $\displaystyle \varpi_{max}=\frac{1}{2}\frac{\lambda+3}{\lambda+2\mu}(\frac{h_B}{r})^2$ is the maximum biaxial strain applied at the nanobubble center, where $h_B$ is the nanobubble maximum height, $\lambda$ and $\mu$ are the Lam\'e coefficients of MoS$_2$ ML~\cite{chirolli}. Note that $\Delta_0^{str}$ only depends on the height-over-radius aspect ratio $\displaystyle \frac{h_B}{r}$, but not on the absolute value of the radius itself. Indeed, for a given substrate (MoS$_2$/substrate), the aspect ratio is universal and thus independent of the nanobubble size \cite{chirolli,Geim}. For MoS$_2$/hBN, an aspect ratio of $0.15$ was observed~\cite{Geim}. Thus, in this case, the strain contribution to $\Delta_0$ is equal to $\Delta_0^{str}=$ $114$ meV.\\

 The dielectric effect on the band gap results by changing locally the dielectric environment through the gap of vacuum between the nanobubble surface and the substrate~\cite{rosner,raja,carmesin}. Hence, the local change of the band gap depends on the interlayer distance, $h$. For MoS$_2$/hBN system, using data from DFT$+GW$ calculations~\cite{florian}, we assume the following interpolation law of the change of $\Delta_0^{env}$ with interlayer distance, 
\begin{equation}
\label{eq22}
\Delta_0^{env}(h_B)= a_0+a_1(1-e^{-a_2(h_B+h_f)^{a_3}})
\end{equation}
 where, $a_0=-0.22585$ eV, $a_1=0.58006$ eV, $a_2=0.21357$ $\angstrom^{-1}$ and $a_3=0.52008$ are fitting parameters. Unlike $\Delta_0^{str}$, since $h_B$ is related to $r$ via the aspect ratio, $\Delta_0^{env}$ depends on the nanobubble radius.\\
\subsubsection{Screened-Coulomb potential} 
 %%%%%%%%%%%%%%%%%%%%%%%%%%%%%%%%%%%%%%%%%%%%%%%%%%%%%%%%%%%%%%%%%%%%%%%%%%%%%%%%%%%%%%%%%%%%%%%%%%%%%%%%%%%%%%%%%%%%%%%%%%
\begin{figure} %[h]
  \begin{center}
     \subfloat[]{
     \includegraphics[width=0.53\textwidth]{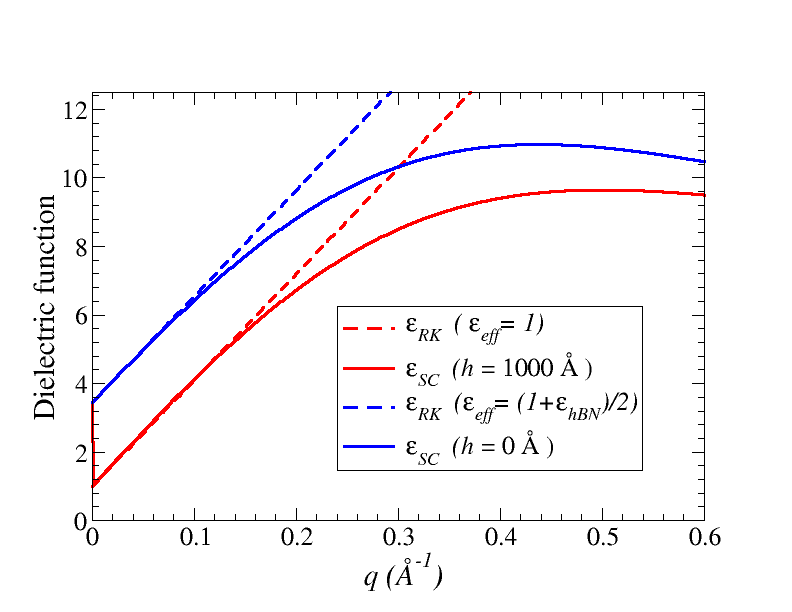}
     \label{fig(11)}
                        }
   \captionsetup{format=plain,labelsep=period}                                                      
   \caption{ (a) Dielectric function $\varepsilon_{SC}(q)$ presented in Eq. (\ref{eq4}) of MoS$_2$ deposed on a hBN substrate of $\varepsilon_{sub}=5.89$~\cite{florian}. The red and blue solid lines represent respectively the cases of an interlayer width of $h=1000$ $\angstrom$ and $h=100$ $\angstrom$. The dashed lines represent the linear behavior of the dielectric function in the Rytova-Keldysh (RK) limit $\varepsilon_{RK}=\varepsilon_{eff}+r^*q$~\cite{keldysh,rytova}.}
    \label{fig(1)}
  \end{center}
\end{figure} 
%%%%%%%%%%%%%%%%%%%%%%%%%%%%%%%%%%%%%%%%%%%%%%%%%%%%%%%%%%%%%%%%%%%%%%%%%%%%%%%%
The Coulomb interaction between particles in MoS$_2$ ML is strongly subject to the surrounding dielectric environment. As it is shown in figure~(\ref{fig(0)}), the field lines of the Coulomb interaction are screened by the adjacent material. In fact, MoS$_2$ ML is separated from the substrate by finite interlayer gap of width $h$~\cite{chirolli,carmesin}. $h$ is equal to $h_f$ and $h_f+h_B$ in the flat region and at the nanobubble centre, respectively. A dielectric environment which takes into account the interlayer gap has been treated in Ref.~\cite{florian} where the screened Coulomb potential $V_{SC}(|\textbf{r}_e-\textbf{r}_h|)$ is given by,
\begin{equation}
\label{eq3}
V_{SC}(|\textbf{r}_e-\textbf{r}_h|)=-e^2\int_{0}^{\infty}dq\frac{J_{0}(|\textbf{r}_e-\textbf{r}_h|q)}{\varepsilon_{SC}(q)}
\end{equation}   
where, $J_0(q|\textbf{r}_e-\textbf{r}_h|)$ is the Bessel function of order zero, $q=\Vert\bm{q}\Vert$ is a two-dimensional wavevector and $\varepsilon(q)$ is the environment-induced nonlocal dielectric function,\\
\begin{equation}
\label{eq4}
\varepsilon_{SC}(q)=\frac{\varepsilon_{3D}(q)(1+\tilde{\varepsilon}_1\tilde{\varepsilon}_2\beta+\tilde{\varepsilon}_1\tilde{\varepsilon}_3\alpha^2\beta+\tilde{\varepsilon}_2\tilde{\varepsilon}_3\alpha^2)}{1+\tilde{\varepsilon}_1\alpha\beta+\tilde{\varepsilon}_2\alpha-\tilde{\varepsilon}_3\alpha+\tilde{\varepsilon}_1\tilde{\varepsilon}_2\beta-\tilde{\varepsilon}_1\tilde{\varepsilon}_3\alpha^2\beta-\tilde{\varepsilon}_2\tilde{\varepsilon}_3\alpha^2-\tilde{\varepsilon}_1\tilde{\varepsilon}_2\tilde{\varepsilon}_3\alpha\beta}
\end{equation}
The different quantities that appear in the above equation, are defined as, $\displaystyle \tilde{\varepsilon}_1=\frac{\varepsilon_g-\varepsilon_{sub}}{\varepsilon_g+\varepsilon_{sub}}$, $\displaystyle  \tilde{\varepsilon}_2=\frac{\varepsilon_{3D}(q)-\varepsilon_g}{\varepsilon_{3D}(q)+\varepsilon_g}$, $\displaystyle \tilde{\varepsilon}_3=\frac{\varepsilon_{t}-\varepsilon_{3D}(q)}{\varepsilon_{t}+\varepsilon_{3D}(q)}$ and $\displaystyle \alpha= e^{-qd}$, $\displaystyle \beta= e^{-q2h}$ where $\varepsilon_t=1$ is the nanobubble top dielectric constant and $\varepsilon_g=1$ the vaccum gap dielectric constant between the surface and the bottom substrate within the nanobubble. $\displaystyle\varepsilon_{3D}(q)$ is the 3D dielectric function of the MoS$_2$~\cite{florian,florian2} and $d$ is the ML thickness. For further need, we define the effective dielectric constant of the bar Coulomb potential, $\varepsilon_{eff}=\frac{\varepsilon_{t}+\varepsilon_{sub}}{2}$.\\

 Comparing the distance $\displaystyle h$ to the wavelength $\displaystyle\frac{1}{q}$, different regimes of screening can be identified. In figure~(\ref{fig(11)}), the dielectric function $\varepsilon_{SC}(q)$ is plotted for a hBN substrate of dielectric constant of $\varepsilon_{sub}=5.89$~\cite{florian}. For the low $q$-values, i.e the long-wavelength regime, the dielectric function show a linear behavior which shifted according to the interlayer width $h$. Here, the dielectric function in Eq.~(\ref{eq4}) can be approximated by:\\
 \begin{equation}
 \varepsilon_{SC}(q)\simeq
 \left\lbrace
 \begin{array}{rcl}
\varepsilon_{RK}(q)=\frac{\varepsilon_{sub}+\varepsilon_{t}}{2}+r^{*}q &\text{for} &qh\simeq0\\
\varepsilon_{RK}(q)=\frac{\varepsilon_{g}+\varepsilon_{t}}{2}+r^{*}q &\text{for} &hq\gg1
 \end{array} 
 \right.
 \end{equation}
 where, $r^{*}$ is the screening length of the free-standing MoS$_2$ ML. In the long-wavelength regime, the linear dependence of $\varepsilon_{SC}(q)$ on $q$ gives rise to the Rytova-Keldysh potential $\varepsilon_{RK}(q)$~\cite{hanan,rytova,keldysh}. This latter was largely adopted to govern the electron-hole interaction mostly in the case of TMD in direct contact to the substrate ($h=0 \angstrom$)~\cite{chernikov,hichri,hichri2,ayari,kylanpaa}. However, according to the Ref.~\cite{hanan}, the Rytova-Keldysh potential approximation, i.e long-wavelength limit approximation, was not considered to be sufficiently accurate to describe the electron-hole interaction in TMD. For this reason and in order to take into account the interlayer gap of width $h$, we use in our work the dielectric function given by Eq.(\ref{eq4}).
\subsection{Solving the Schr\"odinger equation}
Depending on the size of the nanobubble radius versus the average distance between electron and hole ($\text{AD}_{\text{e}-\text{h}}$) in a given state of the unconfined exciton, two limiting cases can help us to resolve the Eq.~(\ref{eq1})~\cite{nanoplatelets,strong,takagahara,koch,bimberg}. Note that what we call here unconfined exciton is an exciton that propagates freely through MoS$_2$ ML, i.e taking $V_{B,i}(r_i)=0$ in Eq.~(\ref{eq1}). In the strong confinement limit~\cite{strong}, the $\text{AD}_{\text{e}-\text{h}}$ of the unconfined exciton states are larger than the nanobubble radius. Here, the wave function shape of the confined exciton is weakly influenced by the Coulomb potential but rather dominated by the confinement potential, the Coulomb potential including only a correction in the lowest eigenenergies. In the weak confinement limit, the $\text{AD}_{\text{e}-\text{h}}$ of the unconfned exciton states are smaller than the radius of the nanobubble, which is usually the case of exciton states with important binding energies~\cite{nanoplatelets}.\\
%%%%%%%%%%%%%%%%%%%%%%%%%%%%%%%%%%%%%%%%%%%%%%%%%%%%%%%%%%%%%%%%%%%%%%%%%%%%%%%%%%%%%%%%%%%%%%%%%%%%%%%%%%%%%%%%%%%%%%%%%%%%%%%%%%%%%%%%%%%%%%%%%
\begin{figure*} %[h]
  \begin{center}
  \subfloat[]{
      \includegraphics[width=0.5\textwidth]{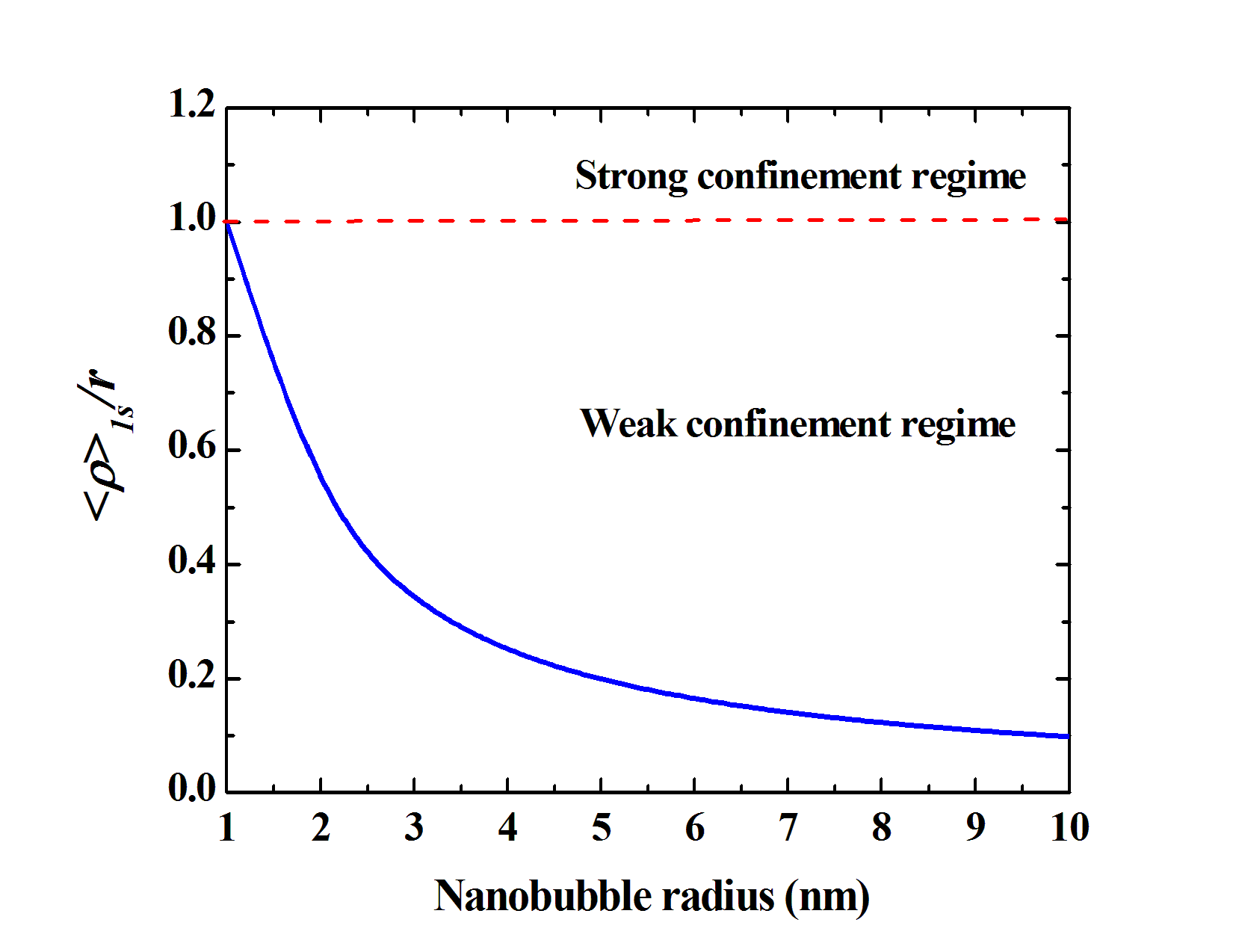}
                         }
   \captionsetup{format=plain,labelsep=period}                                                      
    \caption{The $\tilde{1}s$ average relative distance relatively to the bubble radius $\displaystyle\frac{\displaystyle\braket{\rho}_{\tilde{1}s}}{r}$ versus the nanobubble radius $r$.}
    \label{fig(31)}
  \end{center}
\end{figure*} 
%%%%%%%%%%%%%%%%%%%%%%%%%%%%%%%%%%%%%%%%%%%%%%%%%%%%%%%%%%%%%%%%%%%%%%%%%
%%%%%%%%%%%%%%%%%%%%%%%%%%%%%%%%%%%%%%%%%%%%%%%%%%%%%%%%%%%%%%%%%%%%%%%%%%%%%%%%%%%%%%%%%%%%%%%%%%%%%%%%%%%%%%%%%%%%%%%%%%%%%%%%%%%%%%%%%%%%%%%%%
\begin{figure*} %[h]
  \begin{center}
  \subfloat[]{
      \includegraphics[width=0.5\textwidth]{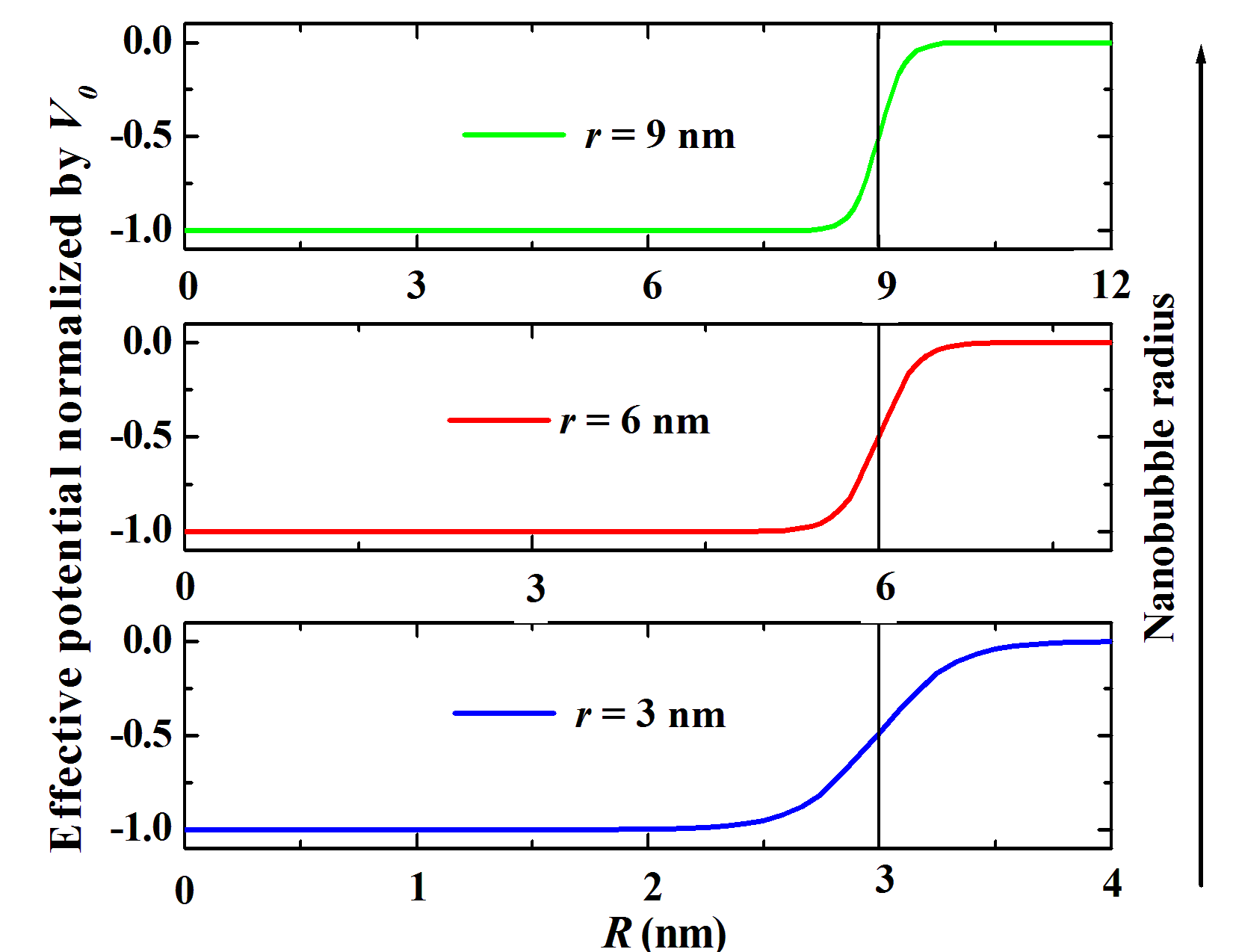}
                         }
   \captionsetup{format=plain,labelsep=period}                                                      
    \caption{The radial effective potentials $\tilde{V}_{B,e}^{\tilde{1}s}(R,\phi)$ of Eq.~(\ref{eq4}) normalized by $V_0$ are shown for three nanobubble radii.}
    \label{fig(32)}
  \end{center}
\end{figure*} 
%%%%%%%%%%%%%%%%%%%%%%%%%%%%%%%%%%%%%%%%%%%%%%%%%%%%%%%%%%%%%%%%%%%%%%%%%
\subsubsection{ Schr\"odinger equation in the weak confinement limit}
 In TMD materials: (i) in the MoS$_2$ ML, the free exciton has an averaged electron-hole separation of the order of few $\angstrom$~\cite{kylanpaa,Berkelbach,Zhang,olsen}. In particular, an averaged electron-hole distance of $10$ $\angstrom$ was found for free-standing MoS$_2$ ML~\cite{kylanpaa,Berkelbach,Zhang,olsen}, and can reach $2$ nm for  environment dielectric constant of, $\displaystyle \varepsilon_{eff}=$ 15~\cite{kylanpaa}; (ii) the radius of (nano)bubbles that were studied either experimentally or theoretically goes from $1$ nm to $1$ $\mu$m~\cite{Geim,chirolli,carmesin}. Therefore, we assume that the confinement region, i.e. the size of the nanobubble, is large enough to confine exciton as a whole, thus placing ourselves in the weak confinement limit. Hence, it is appropriate to pass from electron- and hole-coordinates ($\textbf{r}_e$, $\textbf{r}_h$) to the exciton-coordinates ($\textbf{R}$, $\bm{\rho}$)~\cite{quantum,nanoplatelets}. $\displaystyle\textbf{R}=\frac{m_e\textbf{r}_e+m_h\textbf{r}_h}{m_e+m_h}=(R,\phi)$ is the center of mass (CM) coordinates and  $\bm{\rho}=\textbf{r}_e-\textbf{r}_h=(\rho,\theta)$ is the exciton relative coordinates. The Eq.(\ref{eq1}) then becomes,
\begin{equation}
\label{eq5}
\lbrace-\frac{\hbar^2}{2\mu_X}\nabla_{\rho}^2+V_{SC}(\rho)-\frac{\hbar^2}{2M}\nabla_{R}^2+V_{B}(R,\rho)\rbrace \Psi(\textbf{R},\bm{\rho})=(E_{cf}-E_g)\Psi(\textbf{R},\bm{\rho})
\end{equation}
 where $\displaystyle M=m_e+m_h$ and $\displaystyle \mu_X=\frac{m_em_h}{m_e+m_h}$ represent respectively the total and the reduced effective masses of the exciton.\\
 
Due to the potential term $V_{B}(R,\rho)$ in Eq.~(\ref{eq5}), the centre of mass and relative motion variables are not separable. However, in the weak confinement regime, the relative motion of the exciton is dominated by the screened Coulomb potential and its CM motion is quantized by the confinement potential~\cite{bimberg,koch,nanoplatelets}. Therefore, in order to find a solution to Eq.~(\ref{eq5}), we separate the CM from the relative motion, taking $\displaystyle \Psi(\bm{\rho},\textbf{R})=\psi(\bm{\rho})\Phi(\textbf{R})$ as an approximate trial function~\cite{bimberg,koch,zimmermann1,zimmermann2,nanoplatelets}. Indeed, since electron coordinate, $\displaystyle \textbf{r}_e=\textbf{R}+\frac{1}{1+\sigma}\bm{\rho}$, where $\displaystyle \sigma=\frac{m_e}{m_h}$, the CM and relative motions are correlated via the confinement potential $\displaystyle V_{B}(|\textbf{R}+\frac{1}{1+\sigma}\bm{\rho}|)$.\\

To overcome this difficulty, noticing that in the weak confinement limit the relative motion is not sensitive to potential $V_{B}(R,\rho)$, we will in the following solve first the relative motion Shr\"odinger equation. Then, in order to validate the weak confinement limit approximation, we study the variation in the $\text{AD}_{\text{e}-\text{h}}$ of the free exciton as a function of its dielectric environment. Then, introducing a given obtained eigensolution $\psi(\bm{\rho})$ in $\Psi(\textbf{R},\bm{\rho})$, it will be possible to replace $V_{B}(R,\rho)$ by its partial average $\displaystyle \braket{\psi|V_{B}(R,\rho)|\psi}$ over the relative motion state in equation (\ref{eq5})~\cite{zimmermann1,zimmermann2,nanoplatelets}. Using this procedure, the latter equation becomes separable, and can be solved in turn. Note however that the effective potential we introduced depends on the relative motion state $\psi(\bm{\rho})$.
 \subsubsection{Relative motion solutions}
 We start with the resolution of the relative motion problem in which the relative Schr\"odinger equation is given by, 
\begin{align}
\label{eq6}
\hat{H}_{rel}\psi(\bm{\rho})=\lbrace-\frac{\hbar^2}{2\mu_X}\nabla_{\rho}^2+V_{SC}(\rho)\rbrace\psi(\bm{\rho})&=e\psi(\bm{\rho})
\end{align}
 Here, $\displaystyle V_{SC}(\rho)=-e^2\int dq\frac{J_0(q\rho)}{\varepsilon(q)}$ denote the long-range Coulomb interaction between particles in MoS$_2$ ML screened by the adjacent material. Since the relative motion Hamiltonian $\hat{H}_{rel}$ commutes with the $\hat{L}_z$ angular operator, we can diagonalize $\hat{H}_{rel}$ in the $\hat{L}_z$ Eigen-subspaces, labelled here as $\zeta_l (l\in \mathbb{Z})$. Therefore, it is convenient to expand the relative wave function $\displaystyle \psi(\bm{\rho})$ in a basis of $\displaystyle \phi_{n,l}(\bm{\rho})$ hydrogen-like exact functions, $\displaystyle \psi_{\tilde{n}l}(\bm{\rho})=\sum_{n}^{k}d_{nl}^{\tilde{n}}\phi_{n,l}(\bm{\rho})$. Here, $d_{n,l}^{\tilde{n}}$ are the expansion coefficients, $k$ denotes the basis size and $\displaystyle \phi_{n,l}(\bm{\rho})=N_{n,l} (a_n\rho)^{|l|}L_{n-|l|-1}^{2|l|}(a_n\rho)e^{-\frac{a_{n}}{2}\rho}e^{il\theta}$ are the exact solution of the hydrogenic 2D Schr\"odinger equation for the envelope function. The numbers, $n= 1, 2,...,k$ and $l= 0,\pm 1, \pm 2,..., \pm (n-1)$, denote the radial and the azimuthal quantum numbers of relative motion, respectively. $L_{n-|l|-1}^{2|l|}(a_n\rho)$ are the associated Laguerre polynomial, where $\displaystyle a_n=\frac{4}{(2n-1)a_B}$. $\displaystyle N_{n,l}=\frac{1}{(2\pi)^{1/2}}\lbrace(\frac{2}{(2n-1)a_B})^2\frac{(n-|l|-1)!}{(n+|l|-1)!(2n-1)} \rbrace^{1/2}$ is the normalization constant~\cite{parfitt}. Here, $\displaystyle a_{B}=\frac{\varepsilon_{eff}\hbar^{2}}{\mu_X e^2}$ is the 3D effective Bohr radius.\\

The numerical diagonalization method was adopted for the resolution of Eq.~(\ref{eq6}). In particular, for each subspace of $l$ index,  we perform a numerical diagonalization of the obtained $k\times k$ matrix. Indeed,  for a fixed azimuthal number $l$, we determine the energy spectrum, ( $\displaystyle e_{\tilde{1}l}$, $e_{\tilde{2}l}$,..., $e_{\tilde{n}l}$, ..., $e_{\tilde{k}l}$) and the associated eigenfunctions ( $\psi_{\tilde{1}l}(\bm{\rho})$, $\psi_{\tilde{2}l}(\bm{\rho})$, ..., $\psi_{\tilde{n}l}(\bm{\rho})$, ..., $\psi_{\tilde{k}l}(\bm{\rho})$) where the $\tilde{n}$  index labels the different energy levels. In fact, the index $\tilde{n}$ refer to the dominant contribution of the coefficients $d_{n,l}$ to the excitonic function $\psi_{\tilde{n}l}(\bm{\rho})$, corresponding to the coefficient of the highest weight. Checking the numerical convergence of the calculation when increasing the basis size, we found that a size of $k=8$ states was sufficient to obtain accurate results. We denote $s$ for $l=0$, $p$ for $\displaystyle l=\pm 1$ and $d$ for $\displaystyle l=\pm 2$, that represent the angular symmetries of the relative motion.\\
 \subsubsection{Weak confinement limit model validation}
  Before determining the CM solutions, the validity of the weak confinement assumption should be examined by comparing the $\text{AD}_{\text{e}-\text{h}}$ within the relative motion to the nanobubble radius. In fact, only relative eigenstates with higher binding energies have a $\text{AD}_{\text{e}-\text{h}}$ lower than the nanobubble radius $r$, which support the validity of the weak confinement limit for these states~\cite{nanoplatelets}. Here, the binding energy of the unconfined exciton is defined as $e_{b}^{\tilde{n}s}=\mid e_{\tilde{n}s}\mid$, where $e_{\tilde{n}s}$ denotes the eigenvalue of Eq.~(\ref{eq6}). Based thereon, we choose to restrict ourselves from now on to the ground state $\psi_{\tilde{1}s}(\bm{\rho})$. Furthermore, the $\text{AD}_{\text{e}-\text{h}}$ of the relative $\tilde{1}s$ state given by $\displaystyle\braket{\rho}_{\tilde{1}s}=\braket{\psi_{\tilde{1}s}|\rho|\psi_{\tilde{1}s}}$ depends on both the substrate dielectric constant and the nanobubble hight $\braket{\rho}_{\tilde{1}s}=\braket{\rho}_{\tilde{1}s}(\varepsilon_{sub},h)$. Indeed, this latter is given by $\displaystyle\braket{\rho}_{\tilde{1}s}=\frac{1}{2}\sum_{n}^{k}(3n(n-1)+1)\mid d_{n,0}^{\tilde{1}}\mid^2\times a_{B}$ where the coefficients $d_{n,0}^{\tilde{1}}=d_{n,0}^{\tilde{1}}(\varepsilon_{sub},h)$ and the 3D effective Bohr radius $a_B=a_B(\varepsilon_{sub})$. For MoS$_2$ ML on hBN substrate $\varepsilon_{sub}$ is fixed to 5.89, hence, it remains to determine the interlayer gap width effect on $\displaystyle\braket{\rho}_{\tilde{1}s}$. In fact, the variation of the nanobubble height leads to a variation of the nanobubble radius through the aspect ratio $\frac{h}{r}$. Therefore, $\displaystyle\braket{\rho}_{\tilde{1}s}$ implicitly depends on the bubble radius. To justify the use of the weak confinement limit, we plot in figure~(\ref{fig(31)}) the ratio $\displaystyle\frac{\displaystyle\braket{\rho}_{\tilde{1}s}}{r}$ as a function of $r$.  $\displaystyle\braket{\rho}_{\tilde{1}s}$ decreases rapidly from 1 to much lower values when $r$ increases from 1 nm. As we stated before, the weak confinement limit will take place when the nanobubble radius is greater than the $\text{AD}_{\text{e}-\text{h}}$, $\displaystyle \frac{\braket{\rho}_{\tilde{1}s}}{r}<1$. The latter condition is thus fulfilled for all nanobubble with radius $r$ $>$1 nm.\\
\subsubsection{Center of mass localization in the effective confinement potential}
Turning now to the Eq.~(\ref{eq5}), which in the basis $\Psi(\textbf{R},\bm{\rho})=\psi_{\tilde{1}s}(\bm{\rho})\Phi(\textbf{R})$ becomes,
\begin{equation}
\label{eq7}
\lbrace-\frac{\hbar^2}{2M}\nabla_{\textbf{R}}^2+\tilde{V}_{B}^{\tilde{1}s}(R)\rbrace \Phi_\beta^{\tilde{1}s}(\textbf{R})=E_{\beta}^{\tilde{1}s}\Phi_\beta^{\tilde{1}s}(\textbf{R})
\end{equation}
where, $E_{\beta}^{\tilde{1}s}=E_{cf}^{\tilde{1}s,\beta}-e_{\tilde{1}s}-E_g$ are the CM eigenvalues. As was said above, the solutions of CM motion are not independent of the relative motion solutions. Indeed, the CM eigenfunctions, $\Phi_\beta^{\tilde{1}s}(\textbf{R})$, are influenced by the effective potential, $\displaystyle \tilde{V}_{B}^{\tilde{1}s}(R)$, build from the relative wave function $\psi_{\tilde{1}s}(\bm{\rho})$~\cite{zimmermann1,zimmermann2,nanoplatelets}, 
\begin{equation}
\label{eq8}
\tilde{V}_{B}^{\tilde{1}s}(R)= \int d^2\bm{\rho} \psi_{\tilde{1}s}(\bm{\rho})V_{B}(R,\rho)\psi_{\tilde{1}s}^{*}(\bm{\rho})
\end{equation}
%Note that we took the MoS$_2$ ML band gap as the energy origin, \textit{i.e}, $\displaystyle V_{B}(|\textbf{R}+\frac{1}{1+\sigma}\bm{\rho}|)=-V_0\xi(r-|\textbf{R}+\frac{1}{1+\sigma}\bm{\rho}|)$, see Eq.~(\ref{eq2}).

Numerically found effective potentials per unit of $\Delta_0$ are shown in figure~(\ref{fig(32)}), for nanobubble radii of 3, 6 and 9 nm. According to figure~(\ref{fig(32)}), the relative part affects the confinement potential by inducing small modification at the quantum well barrier; as can be seen, it essentially smoothens the edges of the bubbles potential. This modification becomes more important by decreasing the size of the nanobubble. Therefore, at large bubble radius limit, the confinement potential remains almost unchanged, $\tilde{V}_{B}^{\tilde{1}s}(R)\simeq V_{B}(R)$ which means that the CM motion and the relative motion can be considered as independent. Furthermore, $\tilde{V}_{B}^{\tilde{1}s}(R)$ keeps the cylindrical symmetry of the original confinement potential $V_{B}$. Indeed, there is still an invariance by rotation around the center of the nanobubble. Therefore, the wave function of equation Eq.~(\ref{eq7}) can be expanded in term of Fourier-Bessel series, $\displaystyle\Phi_{\tilde{N}L}^{\tilde{1}s}(\textbf{R})=\sum_{N}^{K}c_{NL}^{\tilde{N}}\chi_{NL}(\textbf{R})$ where $\displaystyle\chi_{NL}(\textbf{R})$ represent the eigenfunctions of a large cylindrical quantum well with infinite barriers and radius $R_c$~\cite{marzin}. The eigenfunctions $\displaystyle\chi_{NL}(\textbf{R})$ are given by $\displaystyle\chi_{NL}(\textbf{R})=\frac{e^{iL\phi}}{\sqrt{\pi} R_c}\frac{J_L(\lambda_{N}^{L}\frac{R}{R_c})}{|J_{L+1}(\lambda_{N}^{L})|}$ where, $\displaystyle\lambda_{N}^{L}$ is the $N^{th}$ zero of the Bessel function $J_L$. The quantum numbers $\displaystyle N\in\mathbb{N^*}$ and $\displaystyle L\in\mathbb{N}$ are the radial and the azimuthal quantum numbers, respectively~\cite{marzin}. We should satisfy the condition, $R_c\gg r$, in order not to perturb the nanobubble eigenstates $\Phi_{\tilde{N}L}^{\tilde{1}s}(\textbf{R})$ by the edges effects. We fix the cylinder radius to $R_c=50$ nm. The CM matrix element is,\\
%%%%%%%%%%%%%%%%%%%%%%%%%%%%%%%%%%%%%%%%%%%%%%%%%%%%%%%%%%%%%%%%%%%%%%%%%%%
\begin{equation}
\label{eq8}
\braket{\chi_{N,L}|H_{R}|\chi_{N',L'}}=\lbrace-\frac{\hbar^2}{2M}(\frac{\lambda^{L}_{N}}{R_c})^2\delta_{NN'}+\tilde{V}_{B, NN'}^{LL'}\rbrace\delta_{LL'}
\end{equation}
where, $\tilde{V}_{B, NN'}^{LL'}=\braket{\chi_{N,L}|\tilde{V}_{B,e}^{\tilde{1}s}(R)|\chi_{N',L'}}$ is calculated numerically. We proceed the same method as the relative motion problem: we determine the energy spectrum via numerical diagonalization of $H_R$ determined for the $\tilde{1}s$ relative motion state. The eigenstates $\Phi_{\tilde{N},L}^{\tilde{1}s}(\textbf{R})$ are labelled $S$ for $L=0$, $P$ for $L= 1$, $D$ for $L= 2$. Here, a basis size of $K=100$  $states$ is found sufficient to give accurate results.\\

For purpose of comparisons, we further use the eigenstates of the unconfined exciton. In fact, the wave function of the unconfined exciton is written in the same way as the confined exciton, a product of the relative and center of mass wave functions. However, in the case of the free exciton the CM wave function is not quantified and written $\displaystyle \Phi(\textbf{R})=\frac{1}{\sqrt{S}}e^{-i\bm{Q.R}}$, where $\bm{Q}$ denotes the in-plane CM wave vector, and $S$ is the normalization area~\cite{hichri,hichri2}.\\

Summarizing this section, we determined the energy of the confined $\tilde{1}s$ exciton in the nanobubble, $\displaystyle E_{cf}^{\tilde{1}s,\tilde{N}L}=E_g+e_{\tilde{1}s}+E_{\tilde{N},L}^{\tilde{1}s}$ and the corresponding eigenfunctions $\Psi_{\tilde{N},L}^{\tilde{1}s}(\textbf{R},\bm{\rho})=\psi_{\tilde{1}s}(\bm{\rho})\displaystyle\Phi_{\tilde{N}L}^{\tilde{1}s}(\textbf{R})$. These functions will be used to determine the oscillator strength as well as the radiative lifetime of the  confined $\tilde{1}s$ exciton. .\\
%%%%%%%%%%%%%%%%%%%%%%%%%%%%%%%%%%%%%%%%%%%%%%%%%%%%%%%%%%%%%%%%%%%%%%%%
\subsection{Oscillator strength and radiative lifetime}
The oscillator strength $f_{\tilde{1}s\tilde{N}L}$ is defined as, $\displaystyle f_{\tilde{1}s\tilde{N}L}=\frac{2}{m_0\hbar \omega_{cf}^{\tilde{1}s,\tilde{N}L}}|\braket{\hat{p}_{\tilde{N}L}^{\tilde{1}s}}|^2$ where $\displaystyle \omega_{cf}^{\tilde{1}s,\tilde{N}L}$ is the optical transition frequency with $\hbar\omega_{cf}^{\tilde{1}s,\tilde{N}L}=E_{cf}^{\tilde{1}s,\tilde{N}L}$. The quantity $\braket{\hat{p}_{\tilde{N}L}^{\tilde{1}s}}$ is the optical matrix element between the crystal ground state and the excited states corresponding to the direct $\tilde{1}s$ exciton confined in the bubble,
\begin{equation}
\label{eq9}
f_{\tilde{1}s\tilde{N}L}=\frac{2|\braket{\bm{u}.\hat{\bm{p}}}_{cv}|^2}{m_0\hbar \omega_{cf}^{\tilde{1}s,\tilde{N}L}}\mid\psi_{\tilde{1}s}(\textbf{0})\mid^2\mid\int \Phi_{\tilde{N}L}^{\tilde{1}s}(\textbf{R})d^2\textbf{R}\mid^2
\end{equation}
where, $\bm{u}$ denotes the photon polarization vector and $\hat{\bm{p}}$ denotes the electron momentum operator. The last factor corresponds to the $K=0$ Fourier transform of the function $\Phi_{\tilde{N}L}^{\tilde{1}s}(\textbf{R})$ we denote $\bar{\Phi}_{\tilde{N}L}^{\tilde{1}s}(\textbf{0})$ in the following.\\

The momentum matrix element $\displaystyle \braket{\bm{u}.\hat{\bm{p}}}_{cv}$ between Bloch functions and the integral over the CM wave function give rise to different kinds of selection rules. In addition, the conditions $\psi_{\tilde{1}s}(\textbf{0})\neq0$ should be satisfied, which is the case only for $\tilde{1}s$ states, as well as $\bar{\Phi}_{\tilde{N}L}^{\tilde{1}s}(\textbf{0})\neq0$. This latter is nonzero only for $L=0$, hence only the $\tilde{1}S$ CM states are optically active. For the matrix element $\displaystyle \braket{\bm{u}.\hat{\bm{p}}}_{cv}=\bm{u}.\braket{\hat{\bm{p}}}_{cv}$, the selection rules come mainly from the coupling term $\braket{\hat{\bm{p}}}_{cv}$ which depend on the Bloch function nature. For bright exciton emission in the direction normal to the nanobubble substrate, the only nonzero elements of the valence-conduction coupling vector lays in the monolayer plane, $\displaystyle \braket{\hat{\bm{p}}_{\pm}}_{cv}=\braket{\frac{\hat{\bm{p}}_{x}\pm i\hat{\bm{p}}_{x}}{\sqrt{2}}}_{cv}$~\cite{Robert,ayari,Wang}. Therefore, only the optical modes with an in-plane polarization components, are non zero. For circularly polarized light, $\sigma_{\pm}$, $\displaystyle \bm{u}_{\pm}=\frac{\bm{u}_x\pm i\bm{u}_y}{\sqrt{2}}$: we obtain $\displaystyle \braket{\bm{u}_{\pm}.\hat{\bm{p}}_{\pm}}_{cv}=\pm\braket{\hat{p}}_{cv}$. The latter can be approximated by means of the $\bm{k.p}$ two band model~\cite{ayari,Wang,Glazov} as $\displaystyle \braket{\hat{p}}_{cv}=\sqrt{\frac{m_0E_p}{2}}$ where $\displaystyle E_p=\frac{m_0E_g}{m_e^*}$ is the Kane energy.\\
Developing the Eq. \ref{eq9}, the oscillator strengths can be written as,
\begin{equation}
\label{eq92}
f_{\tilde{1}s\tilde{N}S}=4\pi\frac{m_0}{m_e^*}\frac{E_g}{\hbar \omega_{cf}^{\tilde{1}s,\tilde{N}L}}|\psi_{\tilde{1}s}(0)|^2R_c^2|\sum_{N}^{K}\frac{c_{N}^{\tilde{N}}}{\lambda_N^0}|^2
\end{equation}

\begin{figure} %[H]
\begin{center}
\subfloat[]{
      \includegraphics[width=0.5\textwidth]{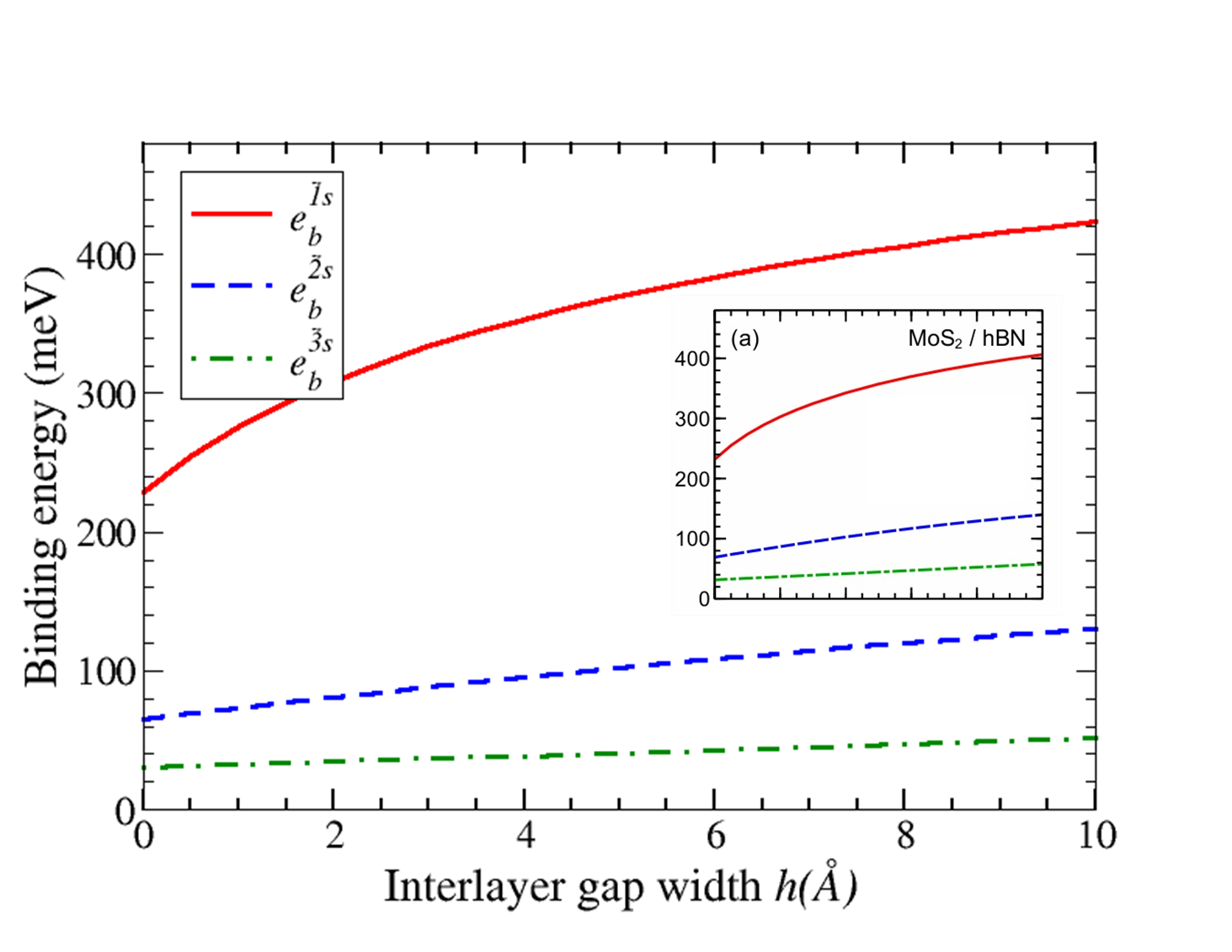}
      \label{fig(201)}
                         }
 \qquad
\subfloat[]{
      \includegraphics[width=0.5\textwidth]{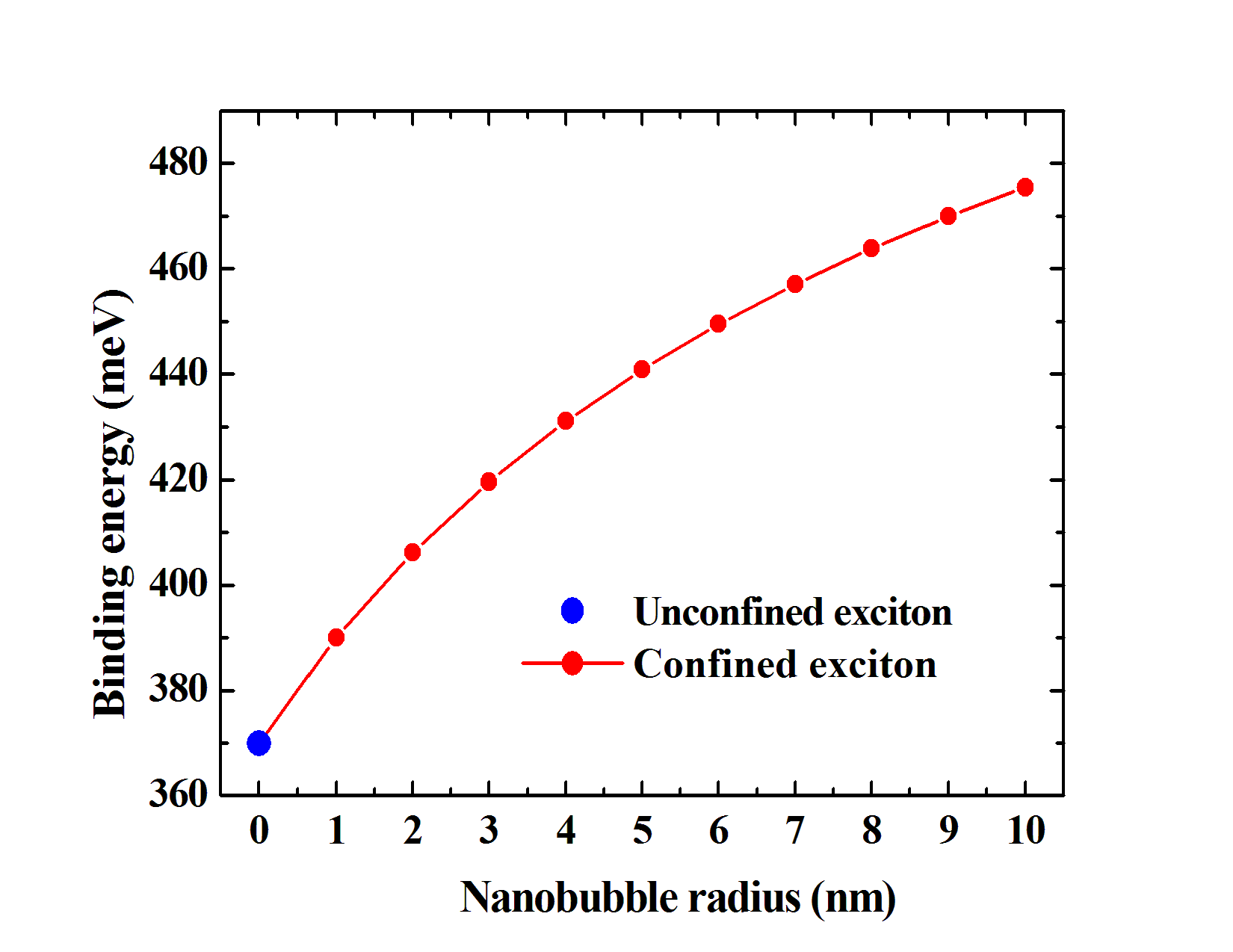}
      \label{fig(202)}
                         }  
\qquad
\subfloat[]{
      \includegraphics[width=0.5\textwidth]{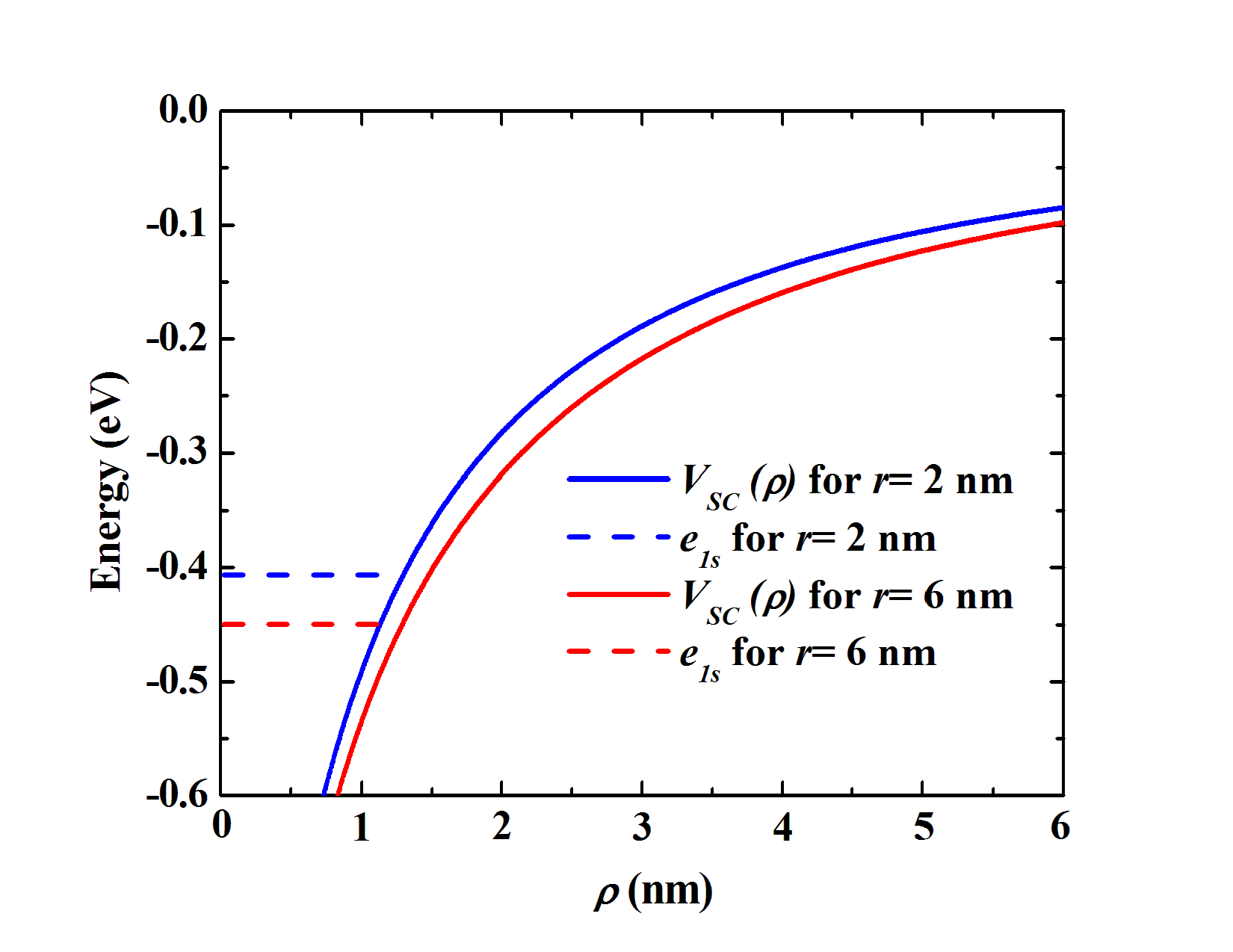}
      \label{fig(203)}
                         }           
\captionsetup{format=plain,labelsep=period}                                                      
\caption{(a) Impact of the interlayer gap on the $\tilde{1}s$, $\tilde{2}s$ and $\tilde{3}s$ exciton binding energies for MoS$_2$ ML on hBN substrate compared to those obtained by Florian \textit{et al.}~\cite{florian} shown in the inset. (b) The $\tilde{1}s$ binding energy of the confined exciton in the nanobubble as a function of the nanobubble radius $r$. The blue circle denotes the $\tilde{1}s$ binding energy of the unconfined exciton in the flat region. (c) The figure presents the screened Coulomb potentials $V_{SC}(\rho)$ of $r=$ 2 nm and $r=$ 6 nm and the corresponding  $\tilde{1}s$ relative energies of the confined exciton.}
    \label{fig(200)}
  \end{center}
\end{figure} 
 %%%%%%%%%%%%%%%%%%%%%%%%%%%%%%%%%%%%%%%%%%%%%%%%%%%%%%%%%%%%%%%
% Furthermore, in order to compare the oscillator strength of the exciton confined in the bubble with that of the unconfined exciton in the flat region $f^{uf}_{\tilde{1}s}$, their ratio,  $\displaystyle\frac{f_{\tilde{1}s\tilde{N}S}}{f_{\tilde{1}s}^{uf}}$, was considered,
 Furthermore, it is more convenient to calculate the normalized oscillator strength of the exciton in the nanobubble by that of the unconfined exciton $f^{uf}_{\tilde{1}s}$ in the flat region with a normalization surface $S=\pi r^2$. Thus, $\displaystyle\frac{f_{\tilde{1}s\tilde{N}S}}{f_{\tilde{1}s}^{uf}}$ reads,
 \begin{equation}
 \label{eq10}
\frac{f_{\tilde{1}s\tilde{N}S}}{f_{\tilde{1}s}^{uf}}=4\frac{\omega_{uf}^{\tilde{1}s}}{\omega_{cf}^{\tilde{1}s,\tilde{N}L}}\frac{\mid\psi_{\tilde{1}s}(0)\mid^2}{\mid\psi_{\tilde{1}s}^{uf}(0)\mid^2}\frac{R_c^2|\sum_{N}^{K}\frac{c_{N}^{\tilde{N}}}{\lambda_N^0}|^2}{r^2}
\end{equation}
where, $\hbar\omega_{uf}^{\tilde{1}s}=E_g+E_{uf}^{\tilde{1}s}$ denotes the exciton energy in the flat region, where $E_{uf}^{\tilde{1}s}=e_{\tilde{1}s}$. Furthermore, radiative lifetime of the confined exciton is inversely proportional to the oscillator strength \cite{ayari},
\begin{equation}
\label{eq11}
\tau_{\tilde{1}s\tilde{N}S}^{-1}=\frac{4n_0e^2}{3m_0\hbar^2c^3}{E_{cf}^{\tilde{1}s,\tilde{N}S}}^2f_{\tilde{1}s\tilde{N}S}
\end{equation}
where, $\displaystyle n_0=\sqrt{\varepsilon_{eff}}$ denotes the effective optical refraction index of the ML environment.
\section{Results and discussion}
\begin{figure} %[h]
\begin{center}
\subfloat[]{
      \includegraphics[width=0.5\textwidth]{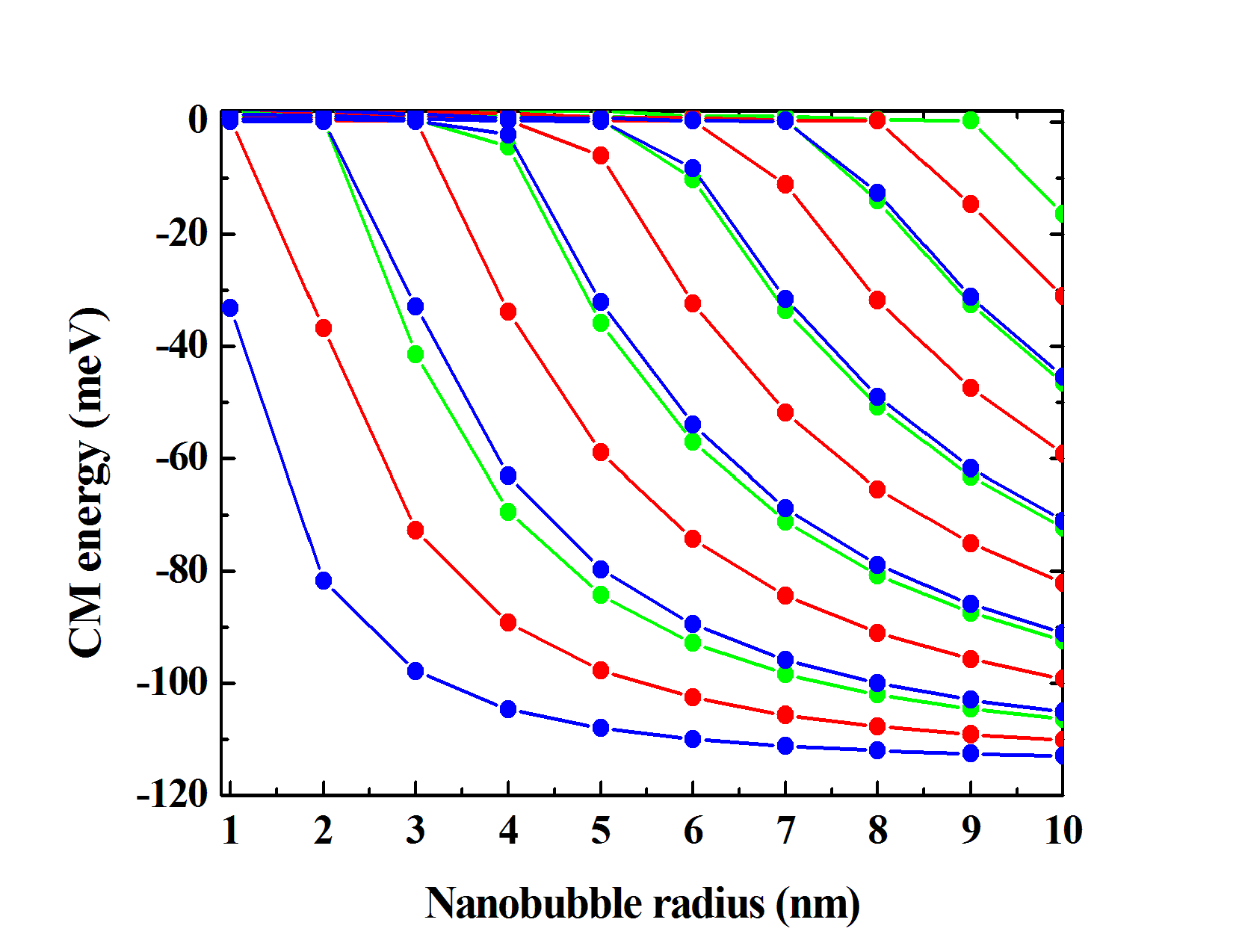}
      \label{fig(204)}
                         } 
\qquad
\subfloat[]{
      \includegraphics[width=0.5\textwidth]{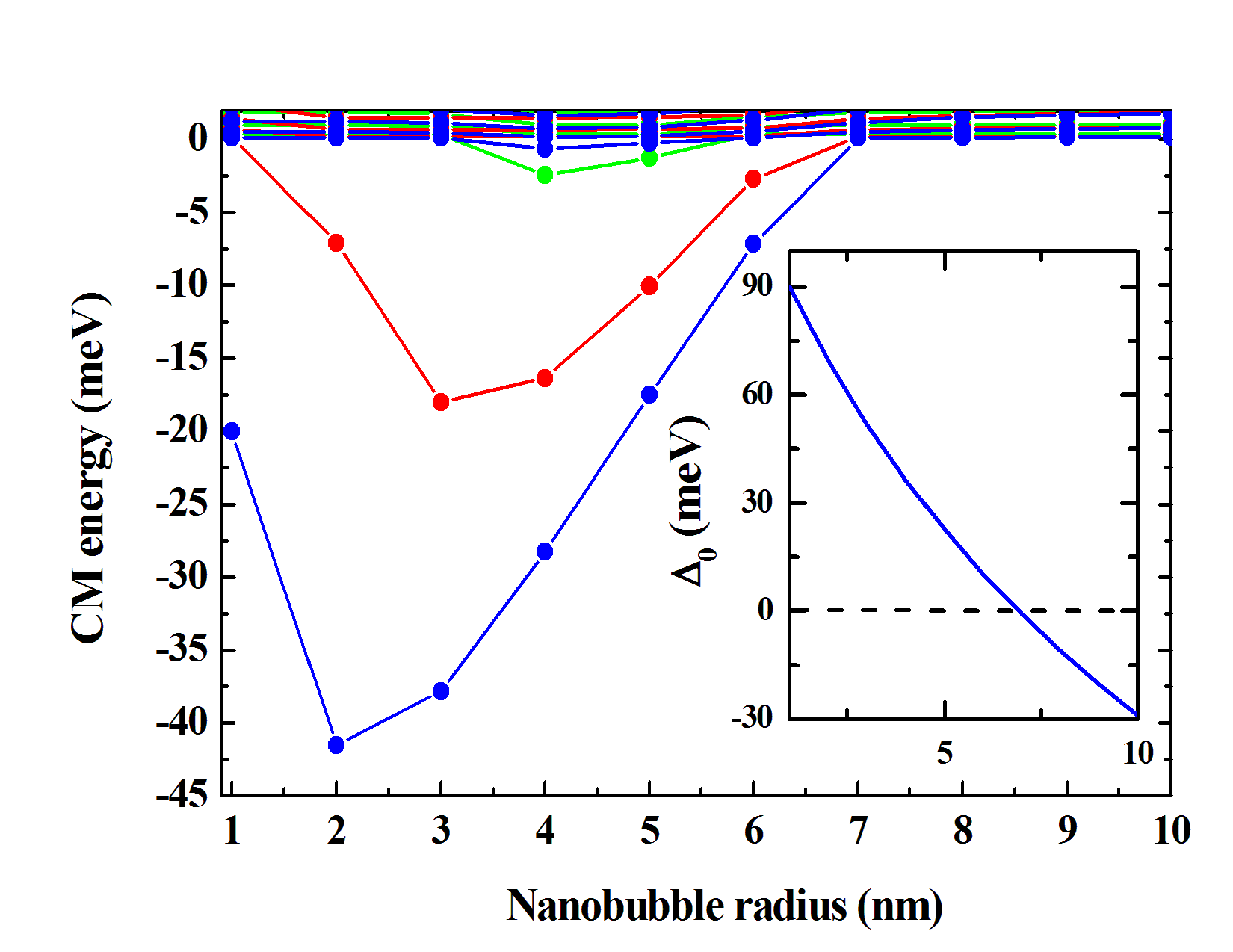}
      \label{fig(205)}
                         } 
\qquad
\subfloat[]{
      \includegraphics[width=0.5\textwidth]{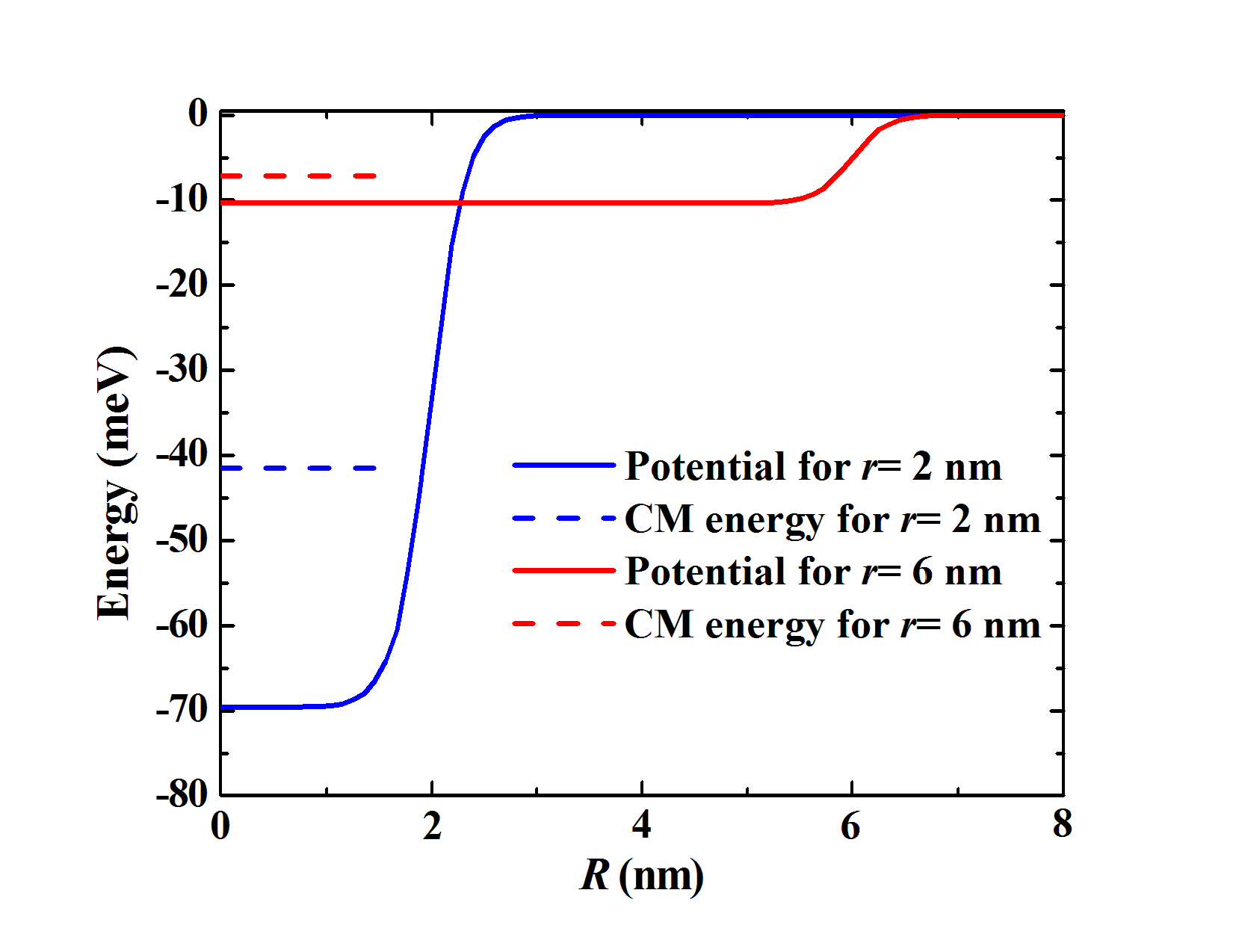}
      \label{fig(2066)}
                         }                 
\captionsetup{format=plain,labelsep=period}                                                      
\caption{Dependence of CM energies on the nanobubble radius $r$. The dielectric induced change in the band offset, $\Delta_{0}^{env}$, is not considered in (a) and is considered in (b). The CM states are represented with: blue lines for $L=0$, red  lines for $L=1$ and green lines for $L=2$; in the inset, the band offset $\Delta_0$. (c) The effective confinement potentials $\tilde{V}_{B,e}^{\tilde{1}s}(R)$ and corresponding $\tilde{1}S$ energy are shown for $r=$ 2 nm and $r=$ 6 nm.}
    \label{fig(206)}
  \end{center}
\end{figure} 

 In this section, we begin by studying separately the exciton relative and CM energies dependence on the nanobubbles radius. Then, we study its effects on the exciton transition energy, oscillator strengths, and radiative lifetimes of the $\tilde{1}s$ confined exciton  in the  nanobubbles of MoS$_2$/hBN system. 
\subsection{Binding energy}
 The exciton relative motion, either in the flat region or in the nanobubble, depends on the interlayer gap width through the screened-Coulomb potential $V_{SC}(\rho)$. This potential was determined by Florian et \textit{al.}~\cite{florian} and used to calculate the exciton binding energies $\displaystyle e_b^{\tilde{1}s}$, $\displaystyle e_b^{\tilde{2}s}$ and $\displaystyle e_b^{\tilde{3}s}$ as a functon of $h$ as shown in the inset of figure~(\ref{fig(201)}). In fact, to calculate $\displaystyle e_b^{\tilde{n}s}$  while the method used by Florian et \textit{al.}~\cite{florian} is a material-realistic description based on the full band structure, our method is based on the effective mass approximation. Therefore, in order to provide confidence in our theoretical model, we reproduce the results of Florian et \textit{al.} in figure~(\ref{fig(201)}). Here, the two excited states of the exciton are taken just for the purpose of comparison. The figure~(\ref{fig(201)}) shows that the binding energies increase by increasing the interlayer gap width. This behavior results from the fact that the electron-hole interaction is less screened by moving away the monolayer from the substrate. Our result is in agreement with that of Florian et \textit{al.}~\cite{florian}. This validates our calculation method used to solve the relative Schr\"odinger of Eq.~(\ref{eq6}).\\

%%%%%%%%%%%%%%%%%%%%%%%%%%%%%%%%%%%%%%%%%%%%%%%%%%%%%%%%%%%%%%%
\begin{figure} %[h]
  \begin{center}
     \subfloat[]{
     \includegraphics[width=0.53\textwidth]{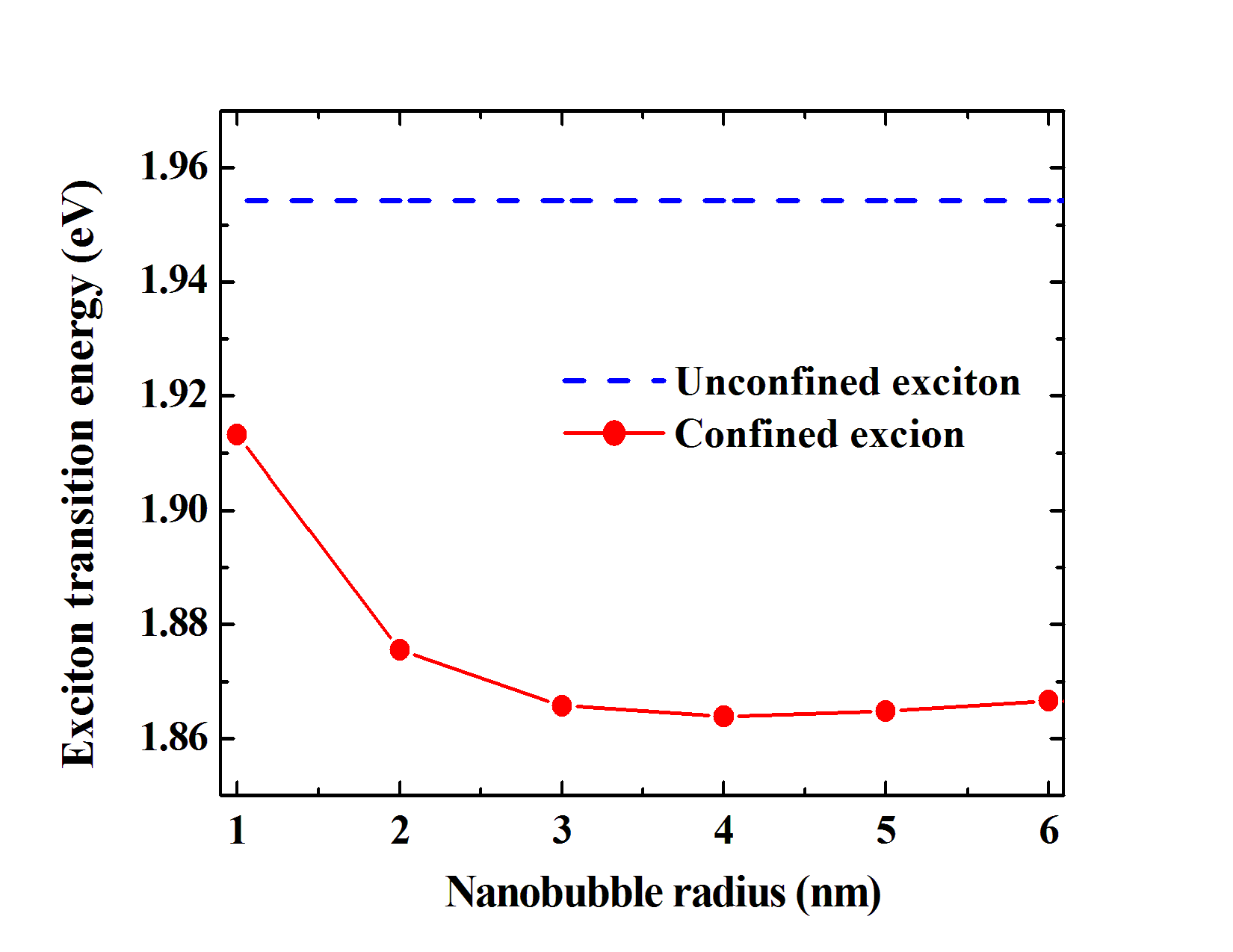}
     \label{fig(51)}
                        }
                   \subfloat[]{
     \includegraphics[width=0.53\textwidth]{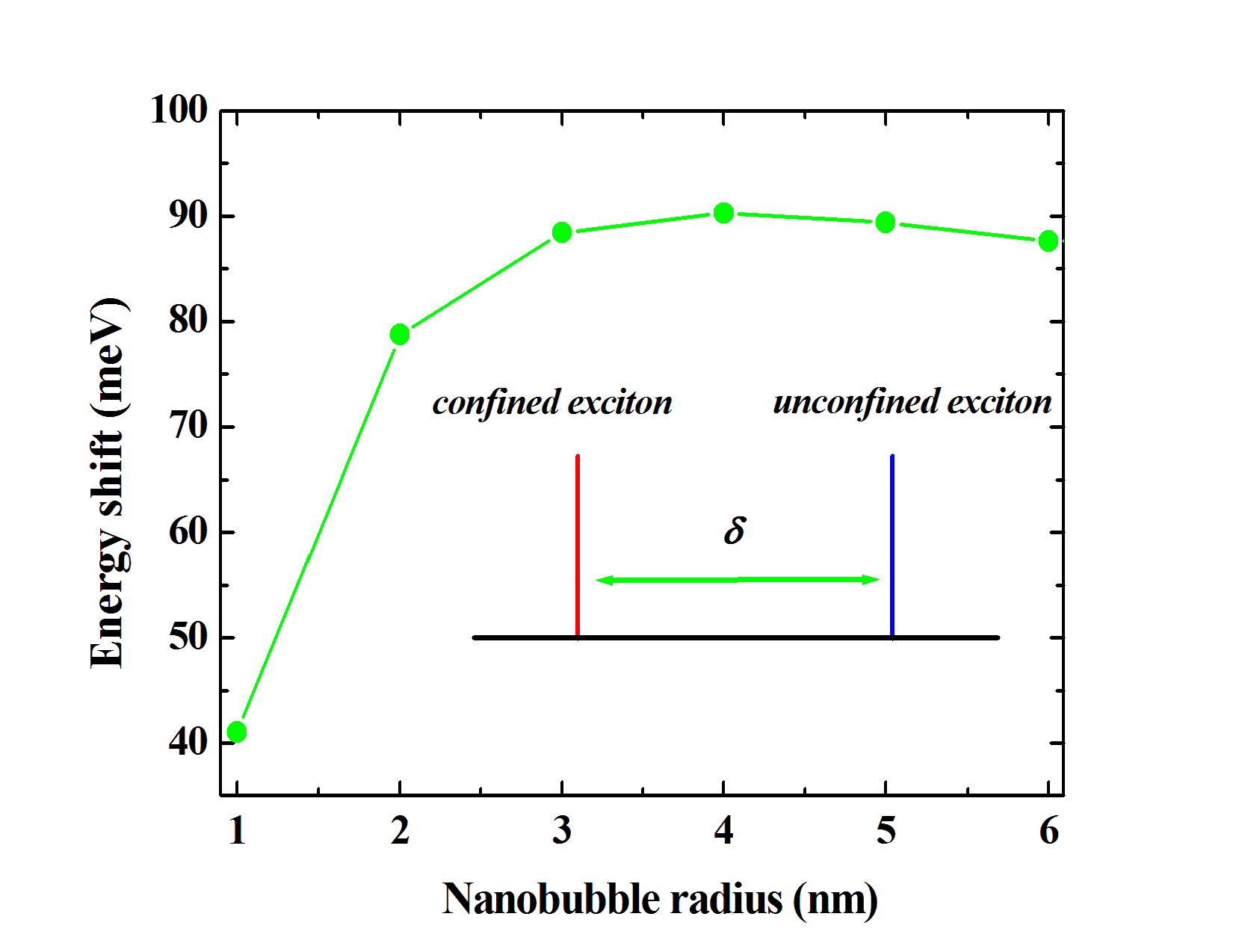}
     \label{fig(52)}
                         } 
   \captionsetup{format=plain,labelsep=period}                                                      
   \caption{ (a) The confined exciton energy $E_{X}^{cf}$, versus the nanobubble radius. (b) The energy difference between the exciton in the flat region and the confined exciton in the nanobubble $\delta$ as a function of $r$.}
    \label{fig(50)}
  \end{center}
\end{figure} 
%%%%%%%%%%%%%%%%%%%%%%%%%%%%%%%%%%%%%%%%%%%%%%%%%%%%%%%%%%%%%%
Furthermore, for the MoS$_2$/hBN flat region, $h_f$ is fixed to $0.5$ nm~\cite{florian} which gives rise to an unconfined exciton binding energy of $\displaystyle e_b^{\tilde{1}s}=369$ meV. However, for the MoS$_2$/hBN nanobubble, the variation of its radius leads to the variation of its hight as they are connected by the aspect ratio $\frac{h_B}{r}=0.15$. Indeed, the distance between the MoS$_2$ nanobubble center and the hBN substrate can be written as a function of the nanobubble radius, $h(nm)=0.5+0.15r$. Therefore, the screened-Coulomb interaction between electron and holes in the nanobubble is expected to be different from that of the flat region and changes with $r$. To get more insight on this behavior, we plot the $\tilde{1}s$ exciton binding energy versus the nanobubbles radius in the figure~(\ref{fig(202)}). In this latter, the $\tilde{1}s$ binding energy of the confined exciton increases with the increase of the nanobubble radius. Therefore, due to dielectric confinement a significant enhancement of the binding energy of the confined exciton with respect to the unconfined one is observed. To explain further, we plot the potential $V_{SC}(\rho)$ for $r=$ 2 nm and $r=$ 6 nm in figure~(\ref{fig(203)}). By increasing the nanobubble size, we see that $V_{SC}(\rho)$ shifts toward lower energy which leads to smaller relative energy. Hence, the enhancement of the binding energy is obtained.\\
\subsection{Exciton CM quantization}
The effects of the nanobubble size on the exciton center of mass quantization are now considered. In fact, $r$ affects not only the size of the quantum well but also its band offset via the dielectric effect on the nanobubble band gap, see Eq.~(\ref{eq22}). Thus, in order to understand of the impact of the dielectric environment on the CM spectrum, we consider two cases with and without the dielectic effect ($\Delta_0^{env}$) as shown in figure~(\ref{fig(206)}). Unlike when $\Delta_0^{env}\neq 0$, $\Delta_0^{env}=0$ stands for the case where the change in $r$ does not affect the energy barrier $\Delta_{0}$. The figure~(\ref{fig(204)}) shows the variation of CM energies, $E_{\tilde{N},L}^{\tilde{1}s}$, as a function of the nanobubble radius for the case where $\Delta_0^{env}$ is omitted. Here, the quantum well depth $\Delta_{0}$ is equal to the strain induced band gap reduction $\Delta_{0}=\Delta_0^{str}$. We find that the number of CM states increases with $r$. In addition, by increasing the nanobubble radius the CM energies decrease. This is due to two effect: (i) the bubble radius $r$ becomes increasingly larger than the relative average distance, $\braket{\rho_{\tilde{1}s}}$, which enhances the localization of exciton CM; (ii) the electron-hole interaction is less screened by moving away the nanobubble center from the substrate, so $\braket{\rho_{\tilde{1}s}}$ decreases. Our results in figure~(\ref{fig(204)}) are quite familiar for quantum dots. Indeed, a comparable behavior of CM energy and the number of states versus $r$ was found by Chirolli et \textit{al.}~\cite{chirolli} for single particle confinement in MoS$_2$ nanobubble.\\
%%%%%%%%%%%%%%%%%%%%%%%%%%%%%%%%%%%%%%%%%%%%%%%%%%%%%%%%%%%%%%%%%%%%%%%%
\begin{figure*} %[h]
  \begin{center}
        \subfloat[]{
      \includegraphics[width=0.45\textwidth]{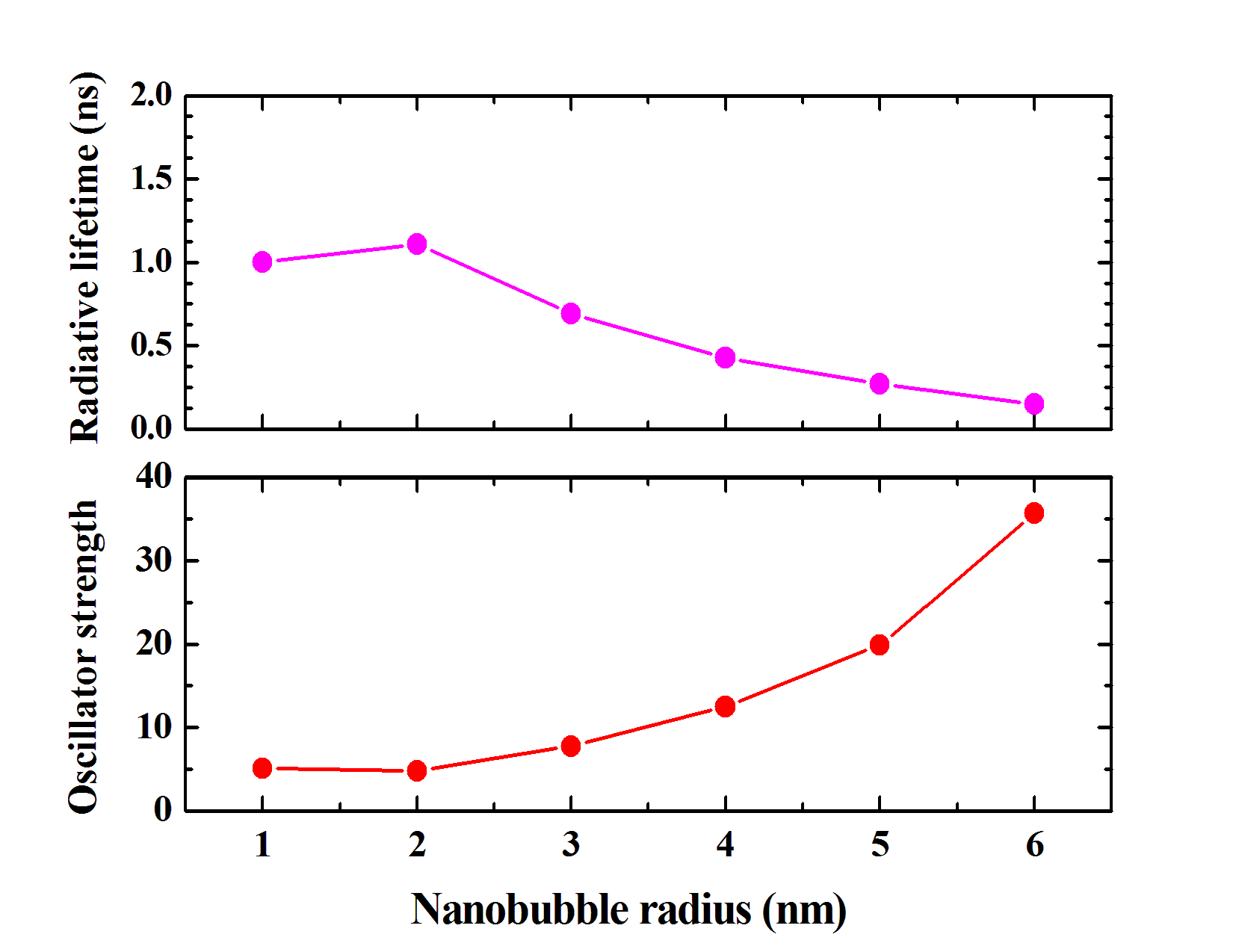}
      \label{fig(61)}
                         } 
       \subfloat[]{
      \includegraphics[width=0.45\textwidth]{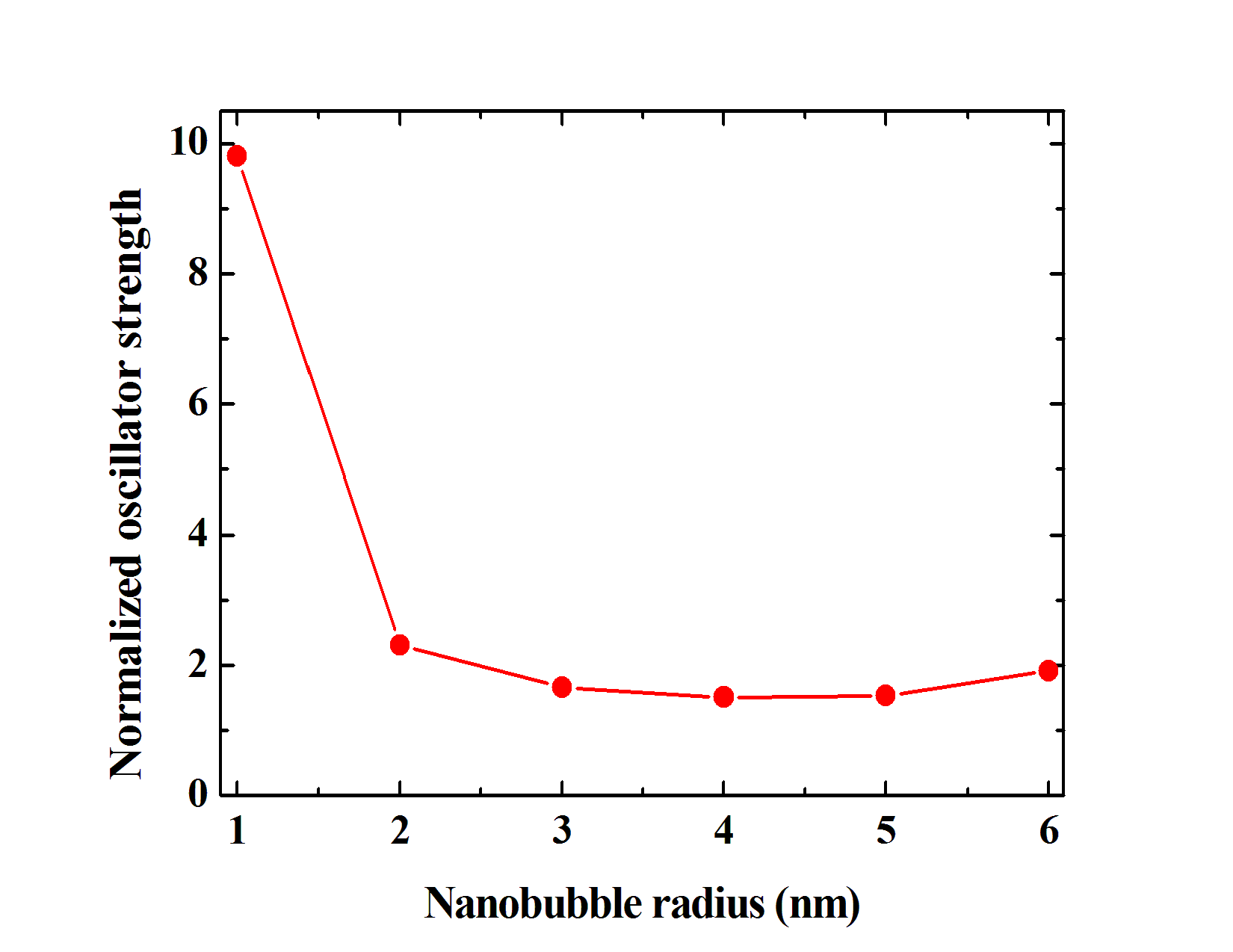}
      \label{fig(62)}
                         } 
       \qquad
       \subfloat[]{
      \includegraphics[width=0.45\textwidth]{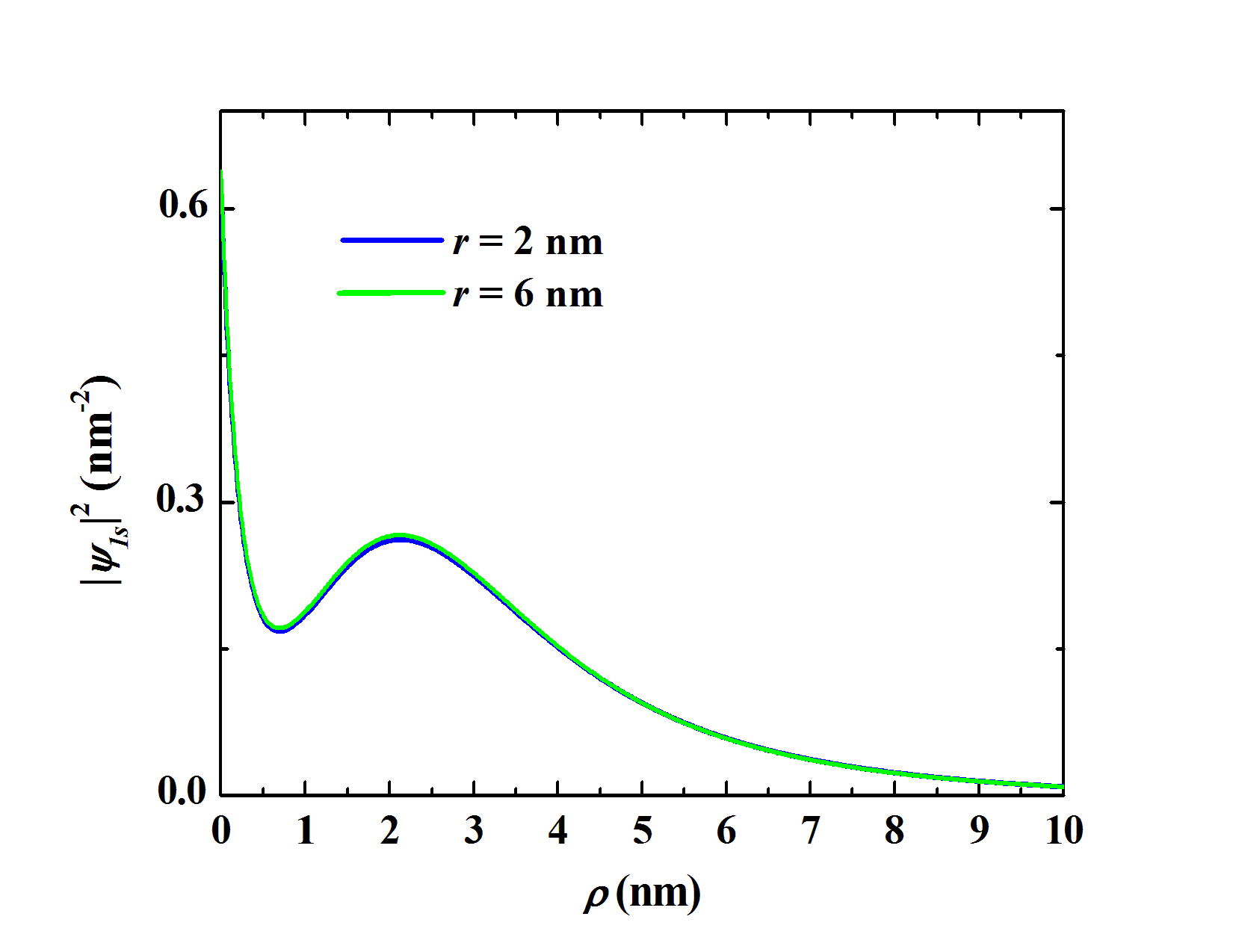}
      \label{fig(63)}
      }
     \subfloat[]{
      \includegraphics[width=0.45\textwidth]{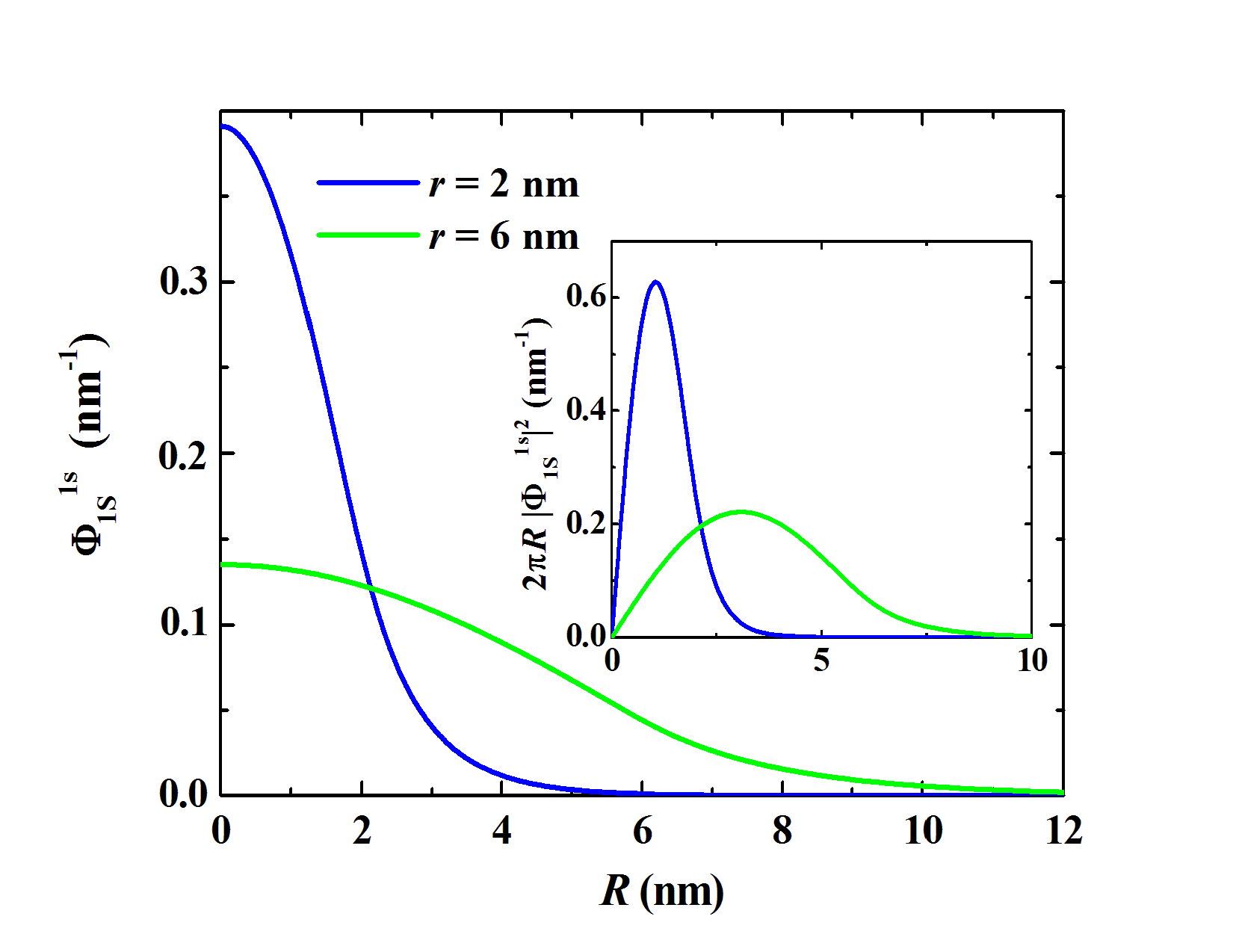}
      \label{fig(64)}
      }
   \captionsetup{format=plain,labelsep=period}                                                      
    \caption{(a) Dependence of oscillator strength (lower panel) and radiative lifetime (upper panel) on the nanobubble radius $r$. (b) The oscillator strength of $\tilde{1}s\tilde{1}S$ state of the confined exciton normalized by the oscillator strength of $\tilde{1}s$ exciton of the flat region as a function of nanobubble radius. The probability densities, $|\psi_{\tilde{1}s}|^2$, of $\tilde{1}s$ relative  state (c) and the wave function of $\tilde{1}S$ CM state (d) are determined for $r=2$ nm and $r=6$ nm. In the inset of (d), the CM radial probability density is plotted.}
    \label{fig(60)}
  \end{center}
\end{figure*} 
%%%%%%%%%%%%%%%%%%%%%%%%%%%%%%%%%%%%%%%%%%%%%%%%%%%%%%%%%%%%%%%

 However, once the dielectric effect included ($\Delta_0^{env}\neq0$), we obtain an unconventional behavior of CM spectrum as presented in figure~(\ref{fig(205)}). The number of the confined states is almost unchanged by increasing the nanobubble size. Indeed, there remains only one state for each orbital symmetry $S$, $P$ and $D$. Since the optically active states must be of symmetry S, the nanobubble of MoS$_2$/hBN can be a single photon source. Besides, the variation of the CM energies as function of $r$ are no longer monotonous. This effect can be explained by the decreasing of $\Delta_{0}$ with $r$ as shown in the inset of figure~(\ref{fig(205)}). Indeed, comparing the confinement effective potentials of $r=2$ nm and $r=6$ nm in figure~(\ref{fig(205)}), it is clear that the band offset, $\Delta_{0}$, drastically decreases while its size increases. Instead of the decreasing with the increasing of the nanobubble size, the energy is forced to increase by the diminution of $\Delta_{0}$.\\
\subsection{Confined exciton energy}
The CM energy and the relative energy present different behavior versus the nanobubble radius. Thus, it is instructive to see the behaviour of their sum in the confined exciton energy as $E_{cf}^{\tilde{1}s\tilde{1}S}=E_g+e_{\tilde{1}s}+E_{\tilde{1}S}^{\tilde{1}s}$. To this end, the nanobubble size effect on $E_{cf}^{\tilde{1}s\tilde{1}S}$ is presented in figure~(\ref{fig(51)}). In the latter, the unconfined exciton energy is also indicated equal to $1.95$ eV. This value is in good agreement with those observed experimentally in Refs.~\cite{mak,mak2}. For the confined exciton, the figure~(\ref{fig(51)}) shows that $E_{cf}^{\tilde{1}s\tilde{1}S}$ decreases with $r$ for $r<$ 3 nm and becomes almost insensitive to $r$ for $r>$ 3 nm. Besides, $E_{cf}^{\tilde{1}s\tilde{1}S}$ is found located around $1.87$ eV. For MoS$_2$/hBN, emitting states was observed in the PL spectrum in a range of energy $1.8$-$1.91$ eV and attributed to localized excitons~\cite{mak,mak2,florian}. Hence, our calculations suggest that the microscopic origin of some of these excitons can be related to the naturally formed nanobubbles. Indeed, Shepard et \textit{al.}~\cite{shepard} have shown that nanobubbles are behind the localized exciton emission in WSe$_2$ ML. Furthermore, in order to compare the energy of the unconfined exciton in the flat region and that of the confined exciton, we plot their difference, $\delta=E_{uf}^{\tilde{1}s}-E_{cf}^{\tilde{1}s\tilde{1}S}$, in figure~(\ref{fig(52)}) as a function of $r$. The increasing of this latter enhances the difference $\delta$ where its energy range is $42$-$90$ meV. These energies are larger than the binding energy of trions which is around 35 meV~\cite{florian}.  Furthermore, our result is comparable to that of quantum dots in WSe$_2$ ML. Indeed, an energy of 20 to 100 meV lower than that of the unconfined exciton was observed for WSe$_2$ quantum dot exciton~\cite{srivastava}.\\
%The $\tilde{1}s\tilde{1}S$ and $\tilde{1}s\tilde{2}S$ oscillator strengths, that are non-vanishing by the selection rules, are shown in figure \ref{fig(51)}.
\subsection{Oscillator strength and radiative lifetime}
Furthermore, we study the influence of nanobubble size on the oscillator strength of the $\tilde{1}s\tilde{1}S$ confined exciton. Hence, in figure~(\ref{fig(61)}), the oscillator strength $f_{\tilde{1}s\tilde{1}S}$ is shown as a function of nanobubble radius. On increasing this latter, the $f_{\tilde{1}s\tilde{1}S}$ increases. This result is in agreement with the dependence found in the Ref.~\cite{que} for quantum dot with disklike shape. The radiative lifetime dependence on $r$ is also studied here. According to the figures~(\ref{fig(61)}), $\tau_{\tilde{1}s\tilde{1}S}$ exhibits the inverse behavior of $f_{\tilde{1}s\tilde{1}S}$, i. e $\tau_{\tilde{1}s\tilde{1}S}$ decreases when $r$ increases. In addition, the radiative lifetime of the confined exciton decreases from 1.1 to 0.15 ns. The resulting radiative lifetimes might be compared with the experimental result for localized exciton of MoS$_2$ ML on a Si/90 nm SiO$_2$ substrate~\cite{lagarde}. Herein, it was found that the radiative lifetime of the localized exciton is equal to 0.125 ns. This value is in good agreement with that of nanobubble radius of the range $3$-$6$ nm. In addition, the calculated radiative lifetime of $\sim 1$ ns is consistent with that found in typical quantum emitters of WSe$_2$ ML~\cite{shepard,srivastava}.\\

To elucidate the origin of $f_{\tilde{1}s\tilde{1}S}$ dependence on $r$, we take the terms which depend on the $r$ contained $f_{\tilde{1}s\tilde{1}S}$ expression given by Eq.~(\ref{eq92}). Three factors determine the magnitude of the oscillator strength as function of $r$; the confined exciton energy, relative wave function and CM wave function. According to figures~(\ref{fig(51)}) and~(\ref{fig(61)}), the variation of the confined exciton energy versus $r$ is not significant compared to that of $f_{\tilde{1}s\tilde{1}S}$. Therefore, one can assume that $E_{cf}^{\tilde{1}s\tilde{1}S}$ is independent of the nanobubble size and not dominate the oscillator strength behavior. The second factor is related to the relative probability density. As shown in figure~(\ref{fig(63)}), the relative probability density are slightly affected by the increase in the nanobubble size. In fact, the increasing of $r$ slightly increases $|\psi_{\tilde{1}s}(\rho)|^2$ intensity.\\
 
 The third factor in determining $f_{\tilde{1}s\tilde{1}S}$ magnitude is related to the CM wave function. Thus, the CM wave function is shown in figure~(\ref{fig(64)}) for $r$ equal to $2$ and $6$ nm. Unlike the relative one, the CM contribution, through $\Phi_{\tilde{1}s}^{\tilde{1}S}(R)$, is considerably affected by increasing $r$. Hence, the change in $f_{\tilde{1}s\tilde{1}S}$ with respect to $r$ is mainly due to the exciton CM localization. The increasing of the nanobubble radius enhances the exciton oscillator strength due to an increase of the area of the CM wave function. This behavior can also be explained in term of radial probability density $2\pi R|\Phi_{\tilde{1}s}^{\tilde{1}S}(R)|^2$, which is shown in the inset of figure~(\ref{fig(64)}). The radial probability density $2\pi R|\Phi_{\tilde{1}s}^{\tilde{1}S}(R)|^2$ extends on a larger surface when the nanobubble size increases which enhances the oscillator strength.\\
 
The previous discussion is true only for the size range of $2$-$6$ nm. In fact, as shown in figure~(\ref{fig(61)}), the oscillator strength (radiative lifetime) has a minimum (maximum) at $r=2$ nm. They have a behavior that corresponds to the cm localization presented in figure~(\ref{fig(205)}). This behavior can be explained by the fact that $\Phi_{\tilde{1}s}^{\tilde{1}S}(R)$ becomes less localized for $r<2$ and the $f_{\tilde{1}s\tilde{1}S}$ increases again with the area occupied by the exciton CM. A similar behavior where the oscillator strength goes to a minimum for certain radius and increases for both the smaller and larger radius, was found for InAs quantum dots~\cite{andreani} and GaAs quantum dots~\cite{hours}.\\

Furthermore, in order to compare the oscillator strength of the confined exciton to that of the unconfined exciton, we normalize $\displaystyle f_{\tilde{1}s\tilde{1}S}$ by $\displaystyle f_{\tilde{1}s}^{uf}$. We plot the normalized oscillator strength against the nanobubble size in figure~(\ref{fig(62)}). It is clear that the ratio $\displaystyle f_{\tilde{1}s\tilde{1}S}/f_{\tilde{1}s}^{uf}$ rapidly when r increases, and then varies much smothly above $r\simeq 3$ nm. In addition, the oscillator strength of the confined exciton is grater than that of the flat region exciton. This result underlines the importance of nanobubble in improving the excitonic oscillator strength. Typically, for WS$_2$ ML, the confined exciton emission dominates the  the PL spectrum which shows the importance of their oscillator forces compared to those of the free exciton~\cite{shepard,srivastava}. This observation is in good agreement with our findings.\\

\section{Conclusion}
In this work, we have studied the properties of exciton confined in the the naturally formed nanobubble of MoS$_2$ ML deposed on hBN. Our calculations show that the optical properties of the nanobubble present a more involved behaviour than that of the free exciton one in monolayers. In particular, an enhancement of the exciton binding energy, oscillator strength and radiative lifetime due to the confinement of exciton and dielectric environment effects have been demonstrated. In addition, only one exciton state per nanobubble was found optically active which is promising for the production of single photon sources. Furthermore, while the increase in the nanobubble size increases the exciton binding energy and oscillator strength, its radiative lifetime decreases. On a radius range of $1$-$6$ nm, the energy of the confined exciton shifts with respect to that of the free exciton by $40$ to $90$ meV. The corresponding radiative lifetime decreases from $1.11$ down to $0.15$ ns, thus reaching values where the localized exciton linewidth should be  strongly dominated by the radiative broadening, opening the way to other interesting applications in optics~\cite{back,Fang,scuri}.
\bibliographystyle{apsrev}
\bibliography{bubble_ref}
\end{document}